\documentclass[12pt]{article}
\usepackage{graphicx}
\usepackage{amsmath}
\usepackage{amssymb}
\usepackage{amsfonts}

\usepackage{times}

\let\OLDthebibliography\thebibliography
\renewcommand\thebibliography[1]{
  \OLDthebibliography{#1}
  \setlength{\parskip}{0pt}
  \setlength{\itemsep}{0pt plus 0.3ex}
}

\newenvironment{sciabstract}{%
\begin{quote} \bf}
{\end{quote}}

\newcounter{lastnote}

\DeclareMathOperator{\Tr}{Tr}
\newcommand{\bra}[1]{\left\langle #1 \right|}
\newcommand{\ket}[1]{\left | #1 \right\rangle}

\newcommand{\mrate}{r_{\mbox{\scriptsize m}}}
\newcommand{\Umw}{U_{\mbox{\scriptsize MW}}}
\newcommand{\Hmw}{H_{\mbox{\scriptsize MW}}}
\newcommand{\Fq}{F_{\mbox{\scriptsize Q}}}

\title{Deterministic Generation of Multiparticle Entanglement by
  Quantum Zeno Dynamics}

\author
{Giovanni Barontini$^1$, Leander Hohmann$^1$, Florian
  Haas$^{1,2}$,\\
J\'er\^ome Est\`eve$^1$ \& Jakob Reichel$^{1\ast}$\\
\\
\normalsize{$^{1}$Laboratoire Kastler Brossel, ENS, UPMC-Paris 6, CNRS,
  Coll\`ege de France,}\\
\normalsize{24 rue Lhomond, 75005 Paris, France}\\
\normalsize{$^{2}$Present address: TWS-Partners AG, 80538 M{\"u}nchen, 
  Germany}\\
\\
\normalsize{$^\ast$To whom correspondence should be addressed; E-mail:
  jakob.reichel@ens.fr}
}

\begin{document} 
\maketitle 

\begin{sciabstract}
{Multiparticle entangled quantum states, a key resource in
  quan\-tum-enhanced metrology and computing, are usually generated by
  coherent operations exclusively. However, powerful new forms of
  quantum dynamics can be obtained when environment coupling is used
  as part of the state generation. Here we demonstrate the use of
  Quantum Zeno Dynamics (QZD), based on non-destructive measurement
  with an optical microcavity, to deterministically generate different
  multi-particle entangled states in an ensemble of 36 qubit atoms in
  less than $5\,\mu$s. We characterize the resulting states by performing
  quantum tomography, yielding a time-resolved account of the
  entanglement generation. We study the dependence on measurement
  strength, and quantify the depth of entanglement. These results show
  that QZD is a versatile tool for fast and deterministic entanglement
  generation in quantum engineering applications. }
\end{sciabstract}

Engineering a desired quantum state -- including the increasingly
complex entangled states required for quantum computing and quantum
simulations -- is usually accomplished using coherent interactions
exclusively, such as a resonant field driving an atomic transition.
Recent developments show that measurement and environment coupling can
also be used as powerful tools for quantum engineering
\cite{Verstraete09,Leroux10,Barreiro11,Krauter11,Shankar13,Lin13,McConnell15}.
One intriguing example is Quantum Zeno Dynamics (QZD), which has been
theoretically studied for more than a decade
\cite{Facchi02,Facchi08,Raimond10}. QZD combines a coherent
interaction with a measurement that is applied simultaneously. In the
simplest case, this measurement detects just a single, initially
unoccupied state in the large state space in which the system
evolves. If the measurement is performed frequently enough,
measurement backaction will keep the detected state unoccupied
indefinitely -- this is the well-known quantum Zeno effect.  However,
the simple fact of blocking one state also entails a profound
modification of the dynamics elsewhere in the state space: this
dynamics, called QZD, remains coherent, but produces states that can
be completely different from those produced by the same coherent
interaction without the measurement
\cite{Facchi02,Facchi08,Raimond10}. Indeed, QZD may
result in quantum states that would be inaccessible in the absence of
measurement, and which are potentially interesting for quantum
engineering. The experimental challenge is to realize the required
non-demolition measurement, which must have high efficiency for
  one state (or a group of states) without affecting the fragile
  quantum coherences in the rest of the state space. Single-particle
QZD has been performed by inducing losses on a specific atomic state
\cite{Schaefer14}. Other recent experiments with single Rydberg atoms
\cite{Signoles14} and photons in a superconducting microwave cavity
\cite{Huard15} have used unitary operators to divide the state
  space and thus obtain an equivalent effect without the use of
measurement.

\begin{figure}[tb]
  \centering
  \includegraphics[height=9cm]{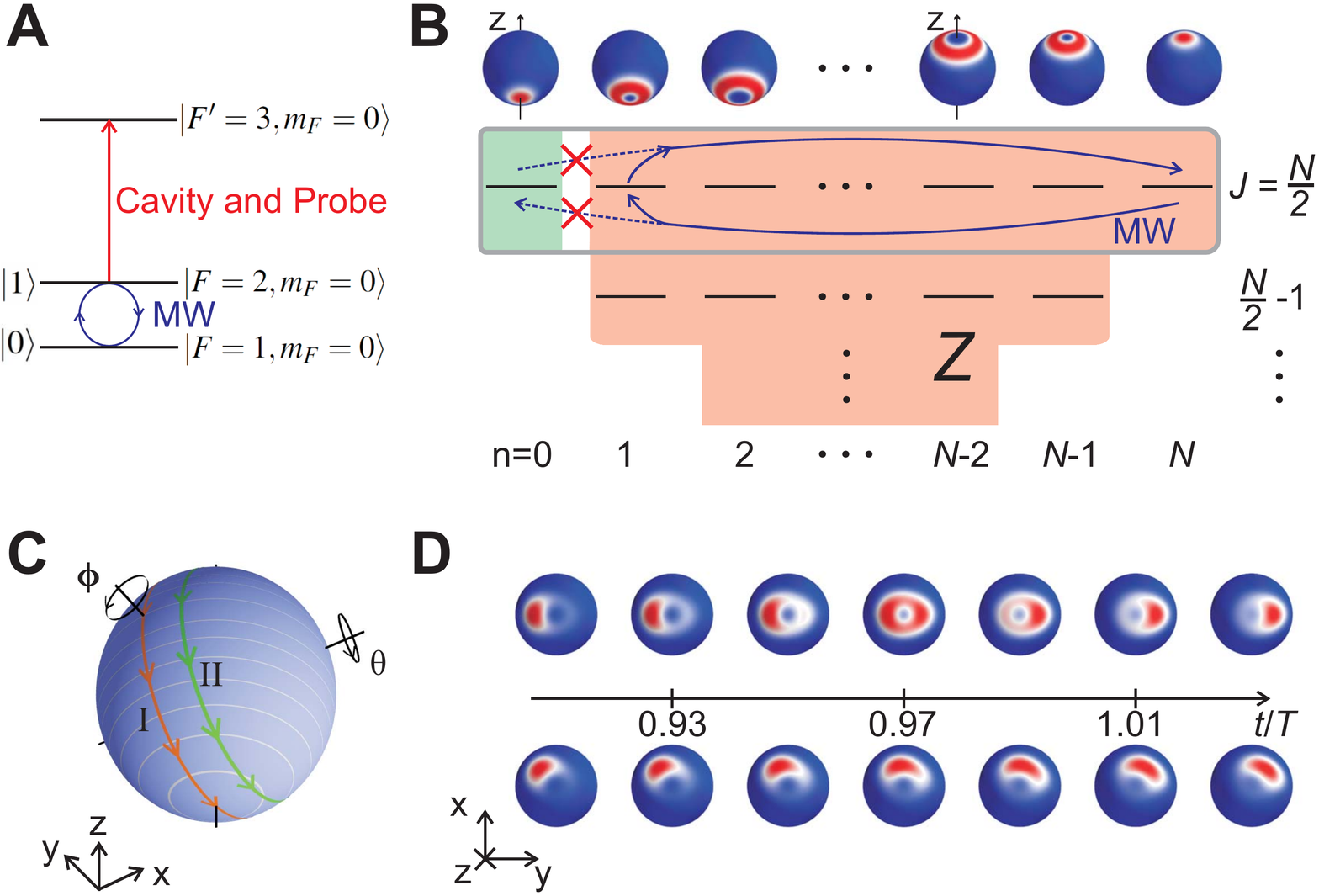}
\caption{\small\textbf{Quantum Zeno dynamics of atoms coupled to a cavity.}
(\textbf{A}) Relevant level scheme of $^{87}$Rb. A resonant 6.8 GHz
microwave allows applying arbitrary rotations to the atomic
qubit. Cavity and probe laser are resonant with the transition
$\ket{1}\to\ket{F'=3,m_{F'}=0}$. The cavity transmits
only when the atoms are in the $\ket{0_N}$ state. (\textbf{B}) The
Hilbert space of the symmetric atomic spin states is spanned by the
Dicke states $\ket{n_N}$ (see text). The Husimi-$Q$ distributions of
some of these states are also displayed. Starting with the atoms in
$\ket{N_N}$, we apply the microwave and simultaneously measure cavity
transmission, leading to coherent evolution that is restricted to the
subspace $Z$ (orange shaded area). (\textbf{C}) Microwave excitations used
in the experiment. Plotted are the Bloch sphere trajectories of the
mean spin without measurement. Trajectory~I drives the mean spin
through the south pole while trajectory~II avoids the pole. (\textbf{D})
Simulated evolution of the Husimi-$Q$ distribution on the Bloch sphere
for 36 atoms on trajectory~I (upper row) and trajectory~II (lower row)
under cavity measurement.}
\end{figure}

Here we use a cavity-based measurement in the regime of high
cooperativity, $C\gg 1$, where $C=g^2/2 \kappa \gamma$, $g$ is the
single-atom coupling rate to the cavity field and $\kappa$ and
$\gamma$ are the atomic and cavity half-linewidths at half
maximum. Compared to other forms of optical detection, the cavity
reduces the spontaneous emission rate by orders of magnitude, making
this a good approximation of an ideal projective measurement
\cite{Volz11,Haas14,MM14}. An ensemble of $^{87}$Rb atoms is confined
in a single antinode of the cavity field, ensuring a near-identical
coupling rate $g= 2\pi\times 190\,$MHz.  Two hyperfine ground state
sublevels serve as qubit states $\ket{0}$ and $\ket{1}$ (Fig.~1A); a
microwave source is used to drive transitions between these states,
realizing a unitary operation $\Umw$. Because the cavity and microwave
fields both provide near-identical coupling to all the atoms, the
relevant Hilbert space consists of the symmetric states of the
collective atomic spin, which is spanned by the Dicke states
$|n_N\rangle$ (Fig.~1B), where $n$ is the number of atoms in $\ket{1}$
and $N$ is the total atom number \cite{MM14}. In our experiment,
cavity and probe beam are tuned as shown in Fig.~1A, so that the
measurement distinguishes the state $\ket{0_N}$ (all atoms in
$\ket{0}$), for which the cavity transmits, from the subspace $Z$ of
all other states, for which it reflects \cite{Haas14}. When $\Umw$
takes the $N$-atom state close to the boundary of $Z$, the dynamics in
presence of the measurement strongly differs from $\Umw$ applied
alone.  The resulting states are highly entangled in general, and
their purity depends on the measurement being strong enough and at the
same time sufficiently nondestructive.  For example, starting from the
initial state $\ket{N_N}$ (all atoms in $\ket{1}$), driving microwave
Rabi oscillations with Rabi frequency $\Omega$ (trajectory~I in
Fig.~1C) would normally produce the state $\ket{0_N}$ after an
evolution time $T=\pi/\Omega$. If the cavity measurement is applied
simultaneously with the microwave drive, however, the system cannot
reach this state. Instead, it evolves through a state which is very
close to the W state, $\ket{1_N}={1}/{\sqrt{N}}(\ket{10\ldots
  0}+\ket{010\ldots0}+ \ldots+\ket{00\ldots 1})$, an
entangled state which is robust against particle loss and enables some
metrological gain over nonentangled states\,\cite{Duer00,Pezze09}. If
instead the microwave drive follows trajectory~II in Fig.~1C, such
that $\Umw$ would go through a state different from but close to
$\ket{0_N}$, then a different entangled state is obtained. Fig.~1D
shows simulated Husimi-$Q$ distributions of the atoms for the two
trajectories in presence of measurement, in the vicinity of the
$\pi$-pulse time.

In the experiment, we use an ensemble of $36\pm 2$ atoms
\cite{MM14}. In order to track the dynamics, we take snapshots of the
atomic state by stopping the microwave and the measurement at
different times $t$ during the evolution. For each $t$, we measure
$Q(\theta,\phi)$ \cite{Haas14,MM14} of the resulting state. In the absence
of measurement, driving the atoms along trajectory~I, we obtain the
usual Rabi oscillation (Fig.~2A, upper panel). By contrast, in the
presence of measurement, we observe that the Husimi-$Q$ distribution is
deformed and features the expected dip in the center (lower panel in
Fig.~2A) when the state reaches the south pole. Figs.~2B and C show
the result of full 2D tomography for different times on trajectories I
and II. In both cases, as the state approaches the boundary of $Z$,
its $Q$ function is deformed such that $Q(0,0)$, which is proportional
to the population in $\ket{0_N}$, remains small at all times. Driving
the dynamics further, the state recovers its gaussian character as it
leaves the measurement boundary.

\begin{figure*}[tb]
  \centering
  \includegraphics[width=0.9\columnwidth]{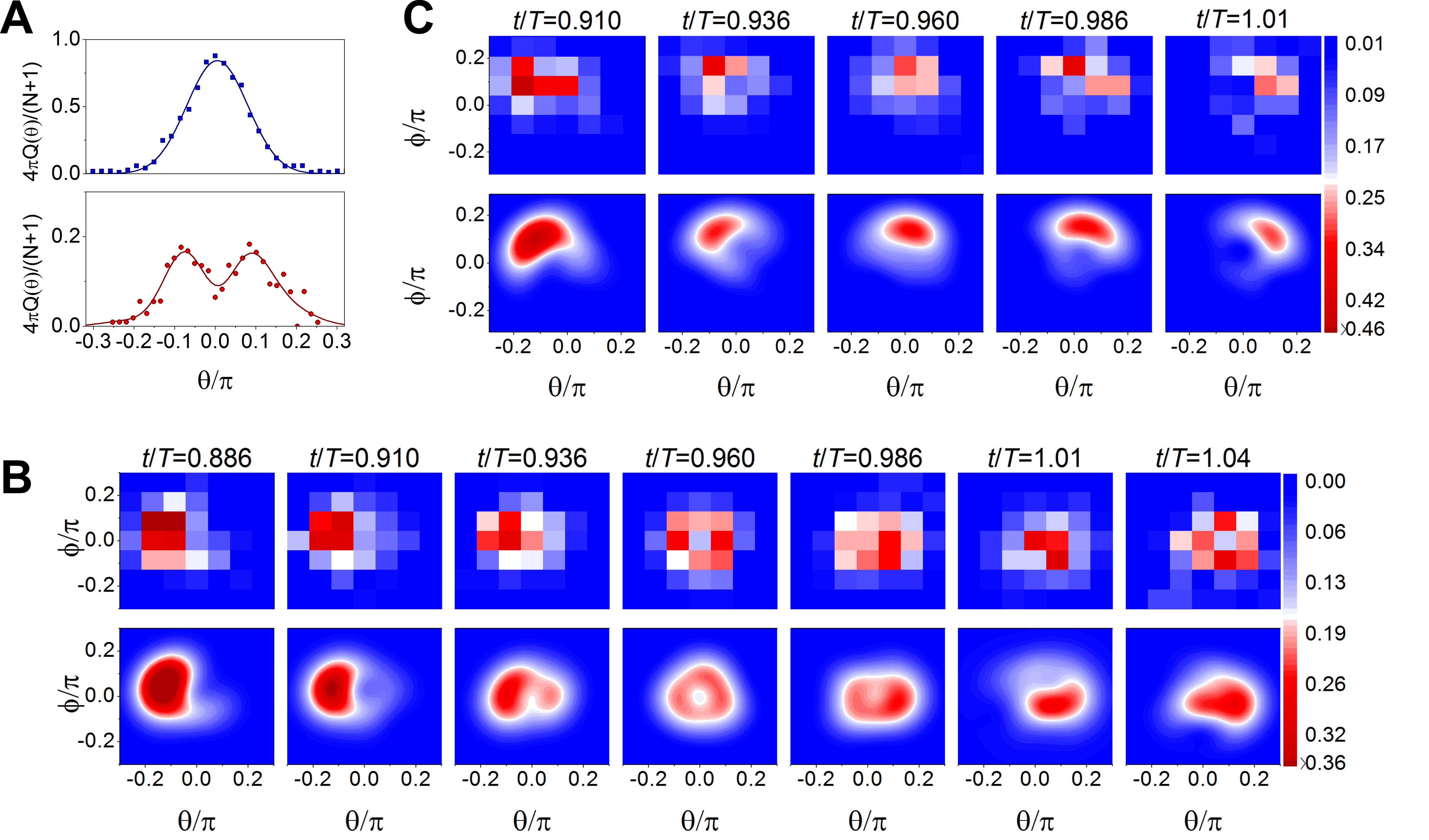}
\caption{\small \textbf{Tomography of quantum Zeno dynamics.} 
(\textbf{A}) Measured high-resolution 1d cuts of the Husimi-$Q$
distributions $Q(\theta,\phi=0)$ when the mean spin reaches the south
pole. The upper panel shows the state after microwave drive along
trajectory~I without measurement. The Husimi-$Q$ function displays the
$\cos(\theta/2)^N$ dependence of the coherent state. The lower panel
results from the same microwave drive, but in the presence of
measurement. The Husimi-$Q$ function shows a characteristic dip. The
lines are maximum-likelihood reconstructions \cite{MM14}. (\textbf{B})
Measured 2d distributions $Q(\theta,\phi)$ at different times for
trajectory~I in presence of measurement. The $7\times 7$ measurement
grid is centered around the south pole. The lower row shows the
Husimi-$Q$ distribution corresponding to the reconstructed density
matrices. For $t/T=0.96$, a ring-shaped distribution appears,
indicating high overlap with the W state. (\textbf{C}) Same as (B),
but for trajectory~II. The quantum Fisher information normalized to
the initial atom number $N=36$ is, from left to right,
$1.51^{+0.39}_{-0.08} 1.25^{+0.3}_{-0.17} 1.47^{+0.14}_{-0.42}
1.19^{+0.25}_{-0.14} 1.14^{+0.25}_{-0.11}$.
The measurement rate is
$\mrate\simeq 22.5\,\Omega$ for all panels. 
}
\end{figure*}

From the tomographic measurements, we reconstruct the symmetric
  part $\rho_s$ of the atomic ensemble density matrix $\rho$ using
a maximum likelihood method \cite{Haas14,MM14}.  On trajectory~I, the
$Q$ function completely encircles the forbidden state for $t=0.96\,
T$, where $T=4.65\,\mu$s.  At this time, the population $\rho_{11}$ of
the W state reaches a maximum of $0.37\pm0.04$, while
$\rho_{00}=0.17\pm0.03$. Knowing $\rho_{11}$ and $\rho_{00}$, we can
use the entanglement depth criterion derived in~\cite{Haas14}: 
any $N$-particle $\rho$ that has a given combination
$\{\rho_{00},\rho_{11}\}$, when decomposed into smaller density
matrices $\rho_1 \otimes \ldots \otimes \rho_M$, must contain at least
one $\rho_i$ with $\dim \rho_i \ge k$, where $k$ depends on the values
$\{\rho_{00},\rho_{11}\}$. Thus, this criterion gives the minimum
number of atoms that are demonstrably entangled with one another, but
makes no statement about the strength of this entanglement, which
might be weak.  Such a criterion of $k$-particle entanglement -- which
does not involve the exact knowledge of $\rho_i$ -- was first derived
for spin-squeezed states \cite{Sorensen01a}. Our
$k(\rho_{00},\rho_{11})$, derived in \cite{Haas14}, efficiently
detects entanglement in the vicinity of the $W$ state. Applied to the
state for $t/T=0.96$, it reveals that this state contains at least
8$^{+3}_{-5}$ entangled particles, in spite of the experimental
imperfections discussed below.

For trajectory~II, the resulting distributions are compressed in one
direction with respect to a coherent state. For such states, our
  entanglement depth criterion is not efficient. To assess their
nonclassical character, we calculate a lower bound on their quantum
Fisher information $\Fq$ \cite{Pezze09,MM14}.   For ideal QZD
  (infinite measurement rate and no loss) on this trajectory,
  numerical simulations indicate a maximum $\Fq/N\simeq 3.4$ reached
  for $t/T\simeq 0.96$. It has been shown
  that $\Fq>N$ is a sufficient (but not necessary) condition for
  entanglement; also, $\Fq/N$ is the maximum achievable reduction in
  the variance of an estimate of a quantum phase using that state,
  with respect to the shot noise limit \cite{Pezze09,Strobel14}.  The
  experimentally realized states in Fig.~2C yield
  $\Fq/N=\{1.51,1.25,1.47,1.19,1.14\}$ \cite{MM14}, showing that
this trajectory too, 
creates states that are entangled and feature some metrological gain
with respect to classical states.

The cavity measurement is characterized by an effective rate which is
not infinitely high, and is accompanied by spontaneous emission, which
tends to populate undesired states outside the symmetric
subspace. Experimentally, the symmetric subspace population
  $\Tr\rho_s$ is readily available from the reconstructed density
  matrices. 
Fig.~3A shows the decay of $\Tr\rho_s$ for trajectory~I. A model
\cite{MM14} including spontaneous emission (solid line), with no
adjustable parameters, reproduces the data well. Apart from this
decay, the features of the QZD can be understood without taking
spontaneous emission into account. Figs.~3B,C show the measured
relative populations $\rho'_{ii}=\rho_{ii}/\Tr\rho_s$ (filled
symbols), comparing them to ideal QZD (dot-dashed lines) and QZD with
finite measurement rate (dashed lines).  $\rho'_{00}$, which would
reach 1 in the absence of measurement, is strongly reduced by the
measurement (Fig.~3B), while $\rho'_{11}$ is increased (Fig.~3C).
We also observe that the turning point of the Rabi oscillation of the
collective spin shifts to shorter times (Fig.~3D), which is expected
because the measurement reduces the dimension of the Hilbert space.
The deviation from the ideal QZD is well described by a simpler model
(dashed lines), which takes into account only the finite measurement
rate, but not the spontaneous emission. The atoms coherently evolve
according to the hamiltonian $\Hmw/\hbar=\Omega J_x$ and are subject to
quantum jumps with a single jump operator
$d=\sqrt{\mrate}\ket{0_N}\bra{0_N}$ accounting for the effect of the
measurement \cite{MM14}. The measurement rate is
$\mrate=2\Phi\sqrt{T_0}$, where $\Phi$ is the photon flux entering the
cavity and $T_0$ the empty-cavity transmission \cite{Volz11}. The
dashed lines are calculated for $\mrate/\Omega=22.5$, the value
expected from the measured photon flux incident onto the cavity, 
$\Phi=21\times 10^6\,\mbox{s}^{-1}$. The
full model including spontaneous emission (solid lines) gives
very similar predictions for $\rho'_{ii}$.

\begin{figure}[tb]
\parbox[c]{.6\linewidth}{\includegraphics[width=\linewidth]{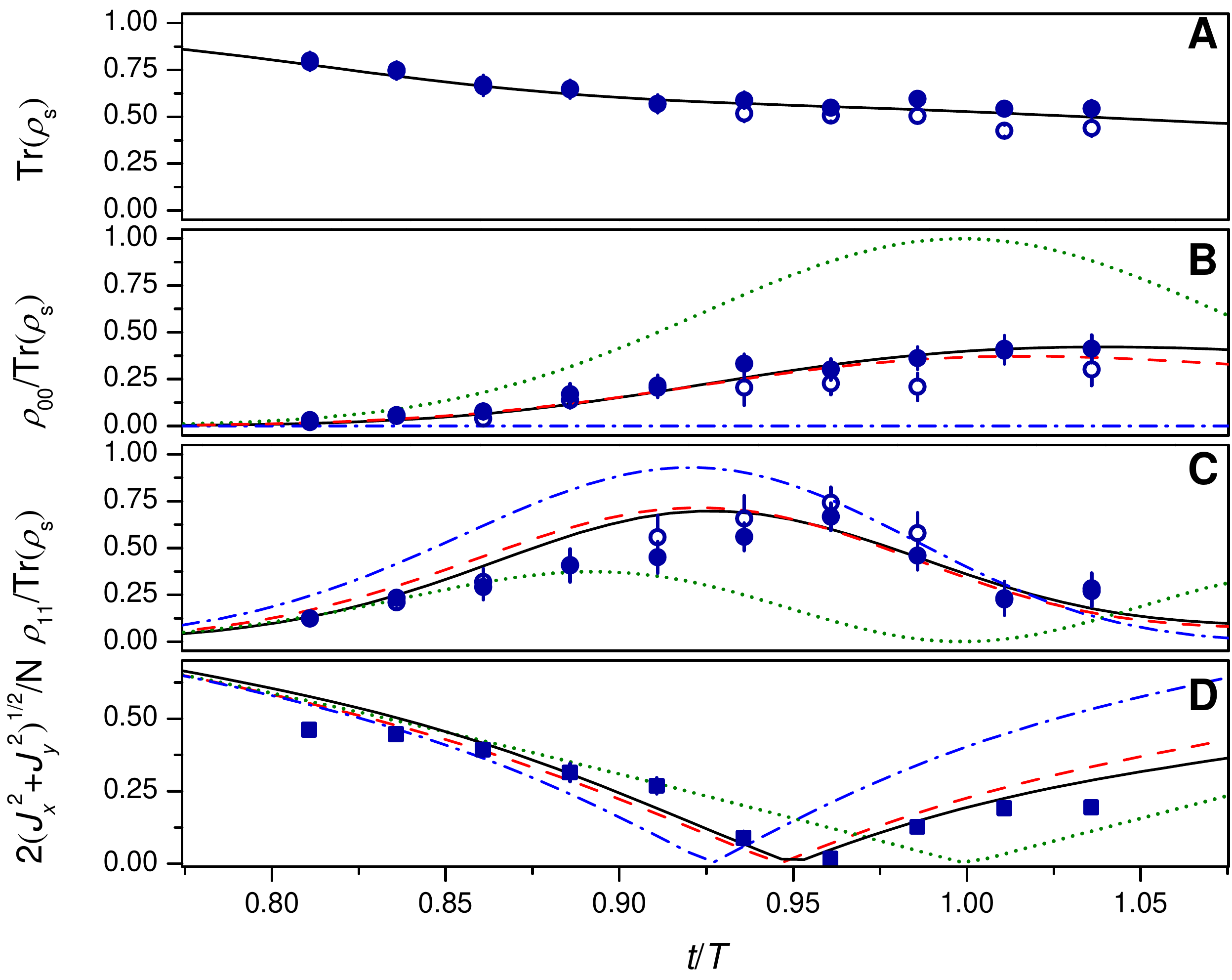}}\hfill%
\parbox[c]{.38\linewidth}{\caption{\small\textbf{Evolution of the Dicke state populations
  during QZD along trajectory~I.}  The measured populations (filled 
symbols) are deduced from the data in Fig.~2B. (\textbf{A}) Population
inside the symmetric subspace. (\textbf{B}), (\textbf{C}) Relative
populations of the Dicke states $\ket{0_N}$ (B) and $\ket{1_N}$
(C). (\textbf{D}) Transverse spin length $2/N\sqrt{J_x^2+J_y^2}$,
where $J_i$ is the $i$-th component of the collective atomic spin.  For
comparison, green dotted lines in (B)-(D) show the expected evolution
without measurement (Rabi oscillation).  The measured data is well
described by a model including spontenous emission \cite{MM14} with no
adjustable parameters
(black solid lines). The dynamics within the symmetric subspace can also be
understood neglecting spontaneous emission (red dashed lines).  Blue
dot-dashed lines are predictions for ideal QZD ($\mrate\to\infty$).
Open symbols: measured evolution
excluding runs with nonzero cavity transmission during the QZD. Error
bars are $1\sigma$ statistical errors of the reconstruction.}
}
\end{figure}

For a given cavity, higher $\Phi$ increases the measurement rate and
thus reduces the contamination of the state by $\ket{0_N}$. However,
it also increases the spontaneous emission rate and thus the
contamination by states outside the symmetric subspace.  The optimum
measurement rate is a compromise between these conflicting effects. We
have investigated this by varying $\Phi$
as shown in Fig.~4.
By solving the full model described above, we obtain the solid curves
in Fig.~4, which are in good agreement with the experimental data, and
which show a broad maximum of $\rho_{11}$ as a function of
$\mrate$. The data in Fig.~2B is taken at $\mrate=22.5\,\Omega$, which
maximizes the number of entangled particles as deduced from the
criterion\,\cite{Haas14}.

By detecting transmitted photons on a photodetector behind the cavity,
we can access the result of the Zeno measurement, i.e., obtain
information about whether the dynamics was indeed restricted to
$Z$. This additional information can be exploited to improve the
fidelity of the produced state, at the price of excluding some
data. In the data of Fig.~2, a transmitted photon is observed for
5-17\% of the runs, depending on $t$. Excluding these runs from the
analysis indeed improves the quality of resulting state, as shown by
the open symbols in Fig.~3 and 4. As an example, for trajectory~I at
$t=0.96\,T$, we obtain an entanglement depth of $11^{+2}_{-3}$ atoms.

\begin{figure}[tb]
\parbox[c]{.6\linewidth}{
		\includegraphics[width=\linewidth]{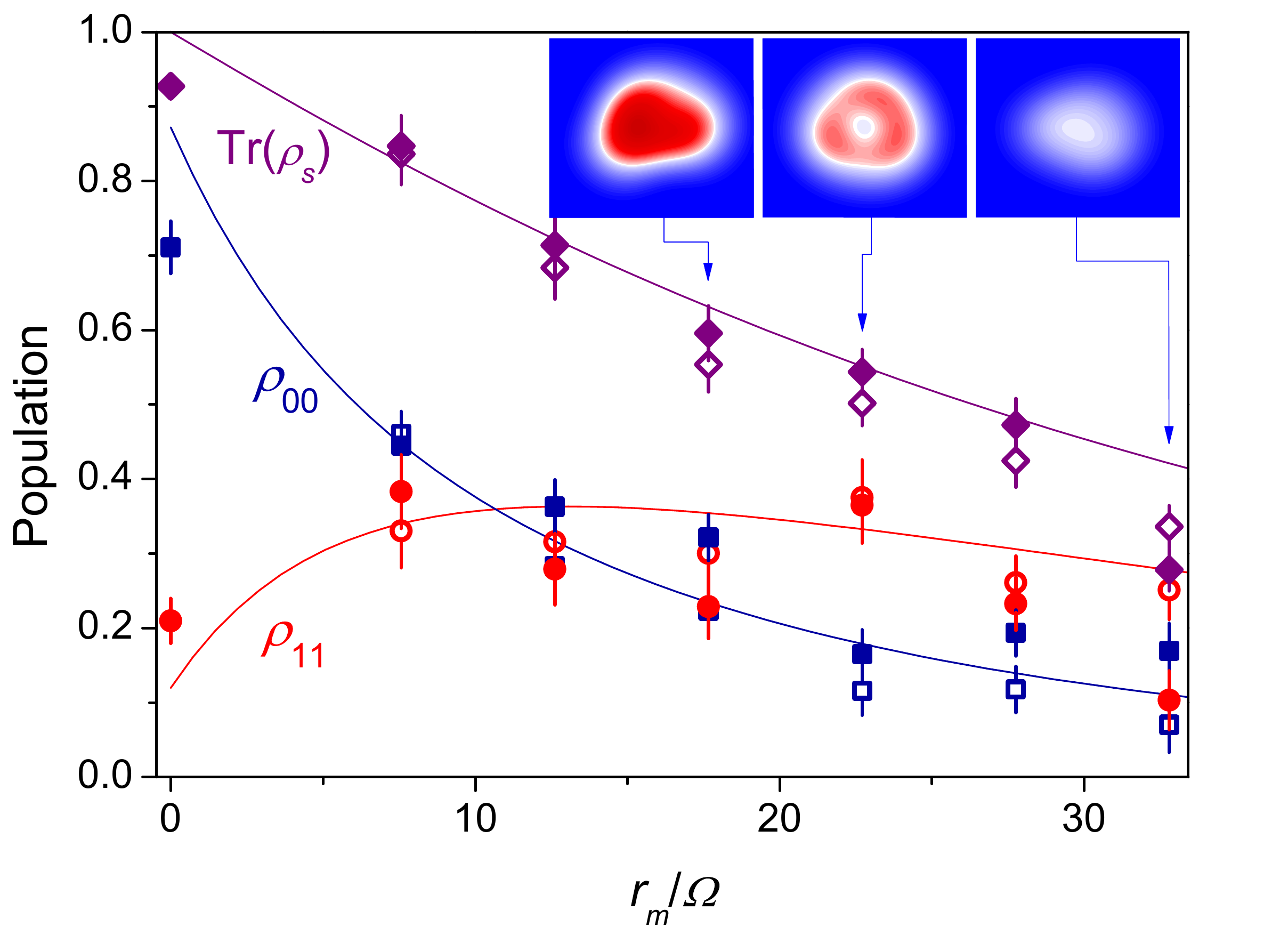}}\hfill%
	\parbox[c]{.38\linewidth}{
		\caption{\small\textbf{QZD for different measurement strengths.}
Tomography measurements are taken after a fixed evolution time
$t=0.96\,T$ for different measurement rates (see text), using trajectory
I. Filled symbols: reconstructed $\rho_{00}$ (blue squares) and
$\rho_{11}$ (red circles), and population in the symmetric subspace
(purple diamonds). Solid lines: results of the full model. The
insets show some of the reconstructed Husimi-$Q$ distributions. Increasing
the effective measurement rate from zero, $\rho_{00}$ decreases while
$\rho_{11}$ increases as normal dynamics turns into QZD. At the same
time, spontaneous emission increases, reducing the population in the
symmetric subspace. Open symbols: runs with nonzero cavity
transmission excluded.}
	}
\end{figure}

The state fidelity can be improved drastically with a better cavity,
using technology that is already available today. In our cavity,
mirror birefringence causes a second, orthogonally polarized TEM00
mode located at a detuning of 540\,MHz from the probed mode. This makes
the spontaneous emission rate much larger than in a birefringence-free
cavity with the same linewidth and $g$.  In the latter, the
spontaneous emission rate for atoms in $\ket{n_N}$ is 
$\mrate/(2nC)$ for $n\ge 1$
\cite{Volz11}.
For our cavity and
transition, the atomic and cavity half-linewidths at half
maximum are $\gamma=2\pi\times 3\,$MHz and $\kappa=2\pi\times
53\,$MHz, so that $C\simeq100$.
Fiber cavities have been fabricated recently with a finesse
approaching 200000 \cite{Muller10,Brandstaetter13,Uphoff15} (instead
of 37000 in our cavity) and with a birefringence reduced to zero
\cite{Uphoff15}. With these improvements alone and all other
parameters unchanged, we calculate a state fidelity $\rho_{11}\simeq
0.8$.

These results demonstrate that QZD is an experimentally feasible tool
for quantum engineering of multiparticle systems. So far, many QZD
proposals have focused on two-qubit systems
\cite{Wang08,Maniscalco08,Rossi09,Shao09,Chandrashekar10,Shi12}. It appears
promising to explore extensions of the scheme studied here. For
example, combining driven Rabi oscillations with a non-demolition
measurement on the equator of the Bloch sphere
\cite{Schleier10,Appel09} could lead to highly entangled states in the
vicinity of the measurement boundary, similar to the mechanism
proposed for photons in \cite{Raimond10}.
\nocite{Kuhr05,Vandersypen05,Gehr10,Rehacek01,Petz11,Toth14,Facchi09}

{
\small
\bibliographystyle{JRWithTitles}
\bibliography{qzd_local}

\begin{thebibliography}{10}
\providecommand{\url}[1]{\texttt{#1}}
\providecommand{\urlprefix}{URL }
\expandafter\ifx\csname urlstyle\endcsname\relax
  \providecommand{\doi}[1]{doi:\discretionary{}{}{}#1}\else
  \providecommand{\doi}{doi:\discretionary{}{}{}\begingroup
  \urlstyle{rm}\Url}\fi
\providecommand{\bibinfo}[2]{#2}
\providecommand{\eprint}[2][]{\url{#2}}

\bibitem{Verstraete09}
\bibinfo{author}{F.~Verstraete}, \bibinfo{author}{M.~M. Wolf}, and
  \bibinfo{author}{J.~I. Cirac}.
\newblock \emph{\bibinfo{title}{Quantum computation and quantum-state
  engineering driven by dissipation}}.
\newblock \bibinfo{journal}{Nat.~Phys.}, \textbf{\bibinfo{volume}{5}},
  \bibinfo{pages}{633} (\bibinfo{year}{2009}).
\newblock \bibinfo{note}{1745-2473}.

\bibitem{Leroux10}
\bibinfo{author}{I.~D. Leroux}, \bibinfo{author}{M.~H. Schleier-Smith}, and
  \bibinfo{author}{V.~Vuleti{\'c}}.
\newblock \emph{\bibinfo{title}{Implementation of Cavity Squeezing of a
  Collective Atomic Spin}}.
\newblock \bibinfo{journal}{Phys.~Rev.~Lett.}, \textbf{\bibinfo{volume}{104}},
  \bibinfo{pages}{073602} (\bibinfo{year}{2010}).

\bibitem{Barreiro11}
\bibinfo{author}{J.~T. Barreiro}, \bibinfo{author}{M.~Muller},
  \bibinfo{author}{P.~Schindler}, \bibinfo{author}{D.~Nigg},
  \bibinfo{author}{T.~Monz}, \bibinfo{author}{M.~Chwalla},
  \bibinfo{author}{M.~Hennrich}, \bibinfo{author}{C.~F. Roos},
  \bibinfo{author}{P.~Zoller}, and \bibinfo{author}{R.~Blatt}.
\newblock \emph{\bibinfo{title}{An open-system quantum simulator with trapped
  ions}}.
\newblock \bibinfo{journal}{Nature}, \textbf{\bibinfo{volume}{470}},
  \bibinfo{pages}{486} (\bibinfo{year}{2011}).
\newblock \bibinfo{note}{0028-0836}.

\bibitem{Krauter11}
\bibinfo{author}{H.~Krauter}, \bibinfo{author}{C.~A. Muschik},
  \bibinfo{author}{K.~Jensen}, \bibinfo{author}{W.~Wasilewski},
  \bibinfo{author}{J.~M. Petersen}, \bibinfo{author}{J.~I. Cirac}, and
  \bibinfo{author}{E.~S. Polzik}.
\newblock \emph{\bibinfo{title}{Entanglement Generated by Dissipation and
  Steady State Entanglement of Two Macroscopic Objects}}.
\newblock \bibinfo{journal}{Phys.~Rev.~Lett.}, \textbf{\bibinfo{volume}{107}},
  \bibinfo{pages}{080503} (\bibinfo{year}{2011}).

\bibitem{Shankar13}
\bibinfo{author}{S.~Shankar}, \bibinfo{author}{M.~Hatridge},
  \bibinfo{author}{Z.~Leghtas}, \bibinfo{author}{K.~M. Sliwa},
  \bibinfo{author}{A.~Narla}, \bibinfo{author}{U.~Vool}, \bibinfo{author}{S.~M.
  Girvin}, \bibinfo{author}{L.~Frunzio}, \bibinfo{author}{M.~Mirrahimi}, and
  \bibinfo{author}{M.~H. Devoret}.
\newblock \emph{\bibinfo{title}{Autonomously stabilized entanglement between
  two superconducting quantum bits}}.
\newblock \bibinfo{journal}{Nature}, \textbf{\bibinfo{volume}{504}},
  \bibinfo{pages}{419} (\bibinfo{year}{2013}).
\newblock \bibinfo{note}{0028-0836}.

\bibitem{Lin13}
\bibinfo{author}{Y.~Lin}, \bibinfo{author}{J.~P. Gaebler},
  \bibinfo{author}{F.~Reiter}, \bibinfo{author}{T.~R. Tan},
  \bibinfo{author}{R.~Bowler}, \bibinfo{author}{A.~S. Sorensen},
  \bibinfo{author}{D.~Leibfried}, and \bibinfo{author}{D.~J. Wineland}.
\newblock \emph{\bibinfo{title}{Dissipative production of a maximally entangled
  steady state of two quantum bits}}.
\newblock \bibinfo{journal}{Nature}, \textbf{\bibinfo{volume}{504}},
  \bibinfo{pages}{415} (\bibinfo{year}{2013}).

\bibitem{McConnell15}
\bibinfo{author}{R.~McConnell}, \bibinfo{author}{H.~Zhang},
  \bibinfo{author}{J.~Hu}, \bibinfo{author}{S.~Cuk}, and
  \bibinfo{author}{V.~Vuleti{\'c}}.
\newblock \emph{\bibinfo{title}{Entanglement with negative Wigner function of
  almost 3000 atoms heralded by one photon}}.
\newblock \bibinfo{journal}{Nature}, \textbf{\bibinfo{volume}{519}},
  \bibinfo{pages}{439} (\bibinfo{year}{2015}).

\bibitem{Facchi02}
\bibinfo{author}{P.~Facchi} and \bibinfo{author}{S.~Pascazio}.
\newblock \emph{\bibinfo{title}{Quantum {Z}eno Subspaces}}.
\newblock \bibinfo{journal}{Phys. Rev. Lett.}, \textbf{\bibinfo{volume}{89}},
  \bibinfo{pages}{080401}.
\newblock \doi{\bibinfo{doi}{10.1103/PhysRevLett.89.080401}}
  (\bibinfo{year}{2002}).

\bibitem{Facchi08}
\bibinfo{author}{P.~Facchi} and \bibinfo{author}{S.~Pascazio}.
\newblock \emph{\bibinfo{title}{Quantum {Z}eno dynamics: mathematical and
  physical aspects}}.
\newblock \bibinfo{journal}{J.~Phys.~A: Math.~Theor.},
  \textbf{\bibinfo{volume}{41}}, \bibinfo{pages}{493001}
  (\bibinfo{year}{2008}).

\bibitem{Raimond10}
\bibinfo{author}{J.~M. Raimond}, \bibinfo{author}{C.~Sayrin},
  \bibinfo{author}{S.~Gleyzes}, \bibinfo{author}{I.~Dotsenko},
  \bibinfo{author}{M.~Brune}, \bibinfo{author}{S.~Haroche},
  \bibinfo{author}{P.~Facchi}, and \bibinfo{author}{S.~Pascazio}.
\newblock \emph{\bibinfo{title}{Phase Space Tweezers for Tailoring Cavity
  Fields by Quantum {Z}eno Dynamics}}.
\newblock \bibinfo{journal}{Phys. Rev. Lett.}, \textbf{\bibinfo{volume}{105}},
  \bibinfo{pages}{213601} (\bibinfo{year}{2010}).

\bibitem{Schaefer14}
\bibinfo{author}{F.~Sch{\"a}fer}, \bibinfo{author}{I.~Herrera},
  \bibinfo{author}{S.~Cherukatti}, \bibinfo{author}{C.~Lovecchio},
  \bibinfo{author}{F.~S. Cataliotti}, \bibinfo{author}{F.~Caruso}, and
  \bibinfo{author}{A.~Smerzi}.
\newblock \emph{\bibinfo{title}{Experimental realization of quantum {Z}eno
  dynamics}}.
\newblock \bibinfo{journal}{Nat.~Commun.}, \textbf{\bibinfo{volume}{5}},
  \bibinfo{pages}{3194} (\bibinfo{year}{2014}).

\bibitem{Signoles14}
\bibinfo{author}{A.~Signoles}, \bibinfo{author}{A.~Facon},
  \bibinfo{author}{D.~Grosso}, \bibinfo{author}{I.~Dotsenko},
  \bibinfo{author}{S.~Haroche}, \bibinfo{author}{J.-M. Raimond},
  \bibinfo{author}{M.~Brune}, and \bibinfo{author}{S.~Gleyzes}.
\newblock \emph{\bibinfo{title}{Confined quantum {Z}eno dynamics of a watched
  atomic arrow}}.
\newblock \bibinfo{journal}{Nat.~Phys.}, \textbf{\bibinfo{volume}{10}},
  \bibinfo{pages}{715} (\bibinfo{year}{2014}).

\bibitem{Huard15}
\bibinfo{author}{L.~Bretheau}, \bibinfo{author}{P.~Campagne-Ibarcq},
  \bibinfo{author}{E.~Flurin}, \bibinfo{author}{F.~Mallet}, and
  \bibinfo{author}{B.~Huard}.
\newblock \emph{\bibinfo{title}{Quantum dynamics of an electromagnetic mode
  that cannot contain N photons}}.
\newblock \bibinfo{journal}{Science}, \textbf{\bibinfo{volume}{348}},
  \bibinfo{pages}{776} (\bibinfo{year}{2015}).

\bibitem{Volz11}
\bibinfo{author}{J.~Volz}, \bibinfo{author}{R.~Gehr},
  \bibinfo{author}{G.~Dubois}, \bibinfo{author}{J.~Est{\`e}ve}, and
  \bibinfo{author}{J.~Reichel}.
\newblock \emph{\bibinfo{title}{Measurement of the internal state of a single
  atom without energy exchange}}.
\newblock \bibinfo{journal}{Nature}, \textbf{\bibinfo{volume}{475}},
  \bibinfo{pages}{210} (\bibinfo{year}{2011}).

\bibitem{Haas14}
\bibinfo{author}{F.~Haas}, \bibinfo{author}{J.~Volz},
  \bibinfo{author}{R.~Gehr}, \bibinfo{author}{J.~Reichel}, and
  \bibinfo{author}{J.~Est{\`e}ve}.
\newblock \emph{\bibinfo{title}{Entangled States of more Than 40 Atoms in an
  Optical Fiber Cavity}}.
\newblock \bibinfo{journal}{Science}, \textbf{\bibinfo{volume}{344}},
  \bibinfo{pages}{180} (\bibinfo{year}{2014}).

\bibitem{MM14}
\bibinfo{note}{See Materials and Methods}.

\bibitem{Duer00}
\bibinfo{author}{W.~D\"ur}, \bibinfo{author}{G.~Vidal}, and
  \bibinfo{author}{J.~I. Cirac}.
\newblock \emph{\bibinfo{title}{Three qubits can be entangled in two
  inequivalent ways}}.
\newblock \bibinfo{journal}{Phys. Rev. A}, \textbf{\bibinfo{volume}{62}},
  \bibinfo{pages}{062314}.
\newblock \doi{\bibinfo{doi}{10.1103/PhysRevA.62.062314}}
  (\bibinfo{year}{2000}).
\newblock \urlprefix\url{http://link.aps.org/doi/10.1103/PhysRevA.62.062314}.

\bibitem{Pezze09}
\bibinfo{author}{L.~Pezze} and \bibinfo{author}{A.~Smerzi}.
\newblock \emph{\bibinfo{title}{Entanglement, Nonlinear Dynamics, and the
  {H}eisenberg Limit}}.
\newblock \bibinfo{journal}{Phys. Rev. Lett.}, \textbf{\bibinfo{volume}{102}},
  \bibinfo{pages}{100401} (\bibinfo{year}{2009}).

\bibitem{Sorensen01a}
\bibinfo{author}{A.~S. S{\o}rensen} and \bibinfo{author}{K.~M{\o}lmer}.
\newblock \emph{\bibinfo{title}{Entanglement and extreme spin squeezing}}.
\newblock \bibinfo{journal}{Phys. Rev. Lett.}, \textbf{\bibinfo{volume}{86}},
  \bibinfo{pages}{4431} (\bibinfo{year}{2001}).

\bibitem{Strobel14}
\bibinfo{author}{H.~Strobel}, \bibinfo{author}{W.~Muessel},
  \bibinfo{author}{D.~Linnemann}, \bibinfo{author}{T.~Zibold},
  \bibinfo{author}{D.~B. Hume}, \bibinfo{author}{L.~Pezz\'e},
  \bibinfo{author}{A.~Smerzi}, and \bibinfo{author}{M.~K. Oberthaler}.
\newblock \emph{\bibinfo{title}{{F}isher information and entanglement of
  non-{G}aussian spin states}}.
\newblock \bibinfo{journal}{Science}, \textbf{\bibinfo{volume}{345}},
  \bibinfo{pages}{424} (\bibinfo{year}{2014}).

\bibitem{Muller10}
\bibinfo{author}{A.~Muller}, \bibinfo{author}{E.~B. Flagg},
  \bibinfo{author}{J.~R. Lawall}, and \bibinfo{author}{G.~S. Solomon}.
\newblock \emph{\bibinfo{title}{Ultrahigh-finesse, low-mode-volume
  {F}abry--{P}erot microcavity}}.
\newblock \bibinfo{journal}{Opt. Lett.}, \textbf{\bibinfo{volume}{35}},
  \bibinfo{pages}{2293}.
\newblock \doi{\bibinfo{doi}{10.1364/OL.35.002293}} (\bibinfo{year}{2010}).

\bibitem{Brandstaetter13}
\bibinfo{author}{B.~Brandst{\"a}tter}, \bibinfo{author}{A.~McClung},
  \bibinfo{author}{K.~Schuppert}, \bibinfo{author}{B.~Casabone},
  \bibinfo{author}{K.~Friebe}, \bibinfo{author}{A.~Stute},
  \bibinfo{author}{P.~O. Schmidt}, \bibinfo{author}{C.~Deutsch},
  \bibinfo{author}{J.~Reichel}, \bibinfo{author}{R.~Blatt}, and
  \bibinfo{author}{T.~E. Northup}.
\newblock \emph{\bibinfo{title}{Integrated fiber-mirror ion trap for strong
  ion-cavity coupling}}.
\newblock \bibinfo{journal}{Rev.~Sci.~Instrum.}, \textbf{\bibinfo{volume}{84}},
  \bibinfo{pages}{123104} (\bibinfo{year}{2013}).

\bibitem{Uphoff15}
\bibinfo{author}{M.~Uphoff}, \bibinfo{author}{M.~Brekenfeld},
  \bibinfo{author}{G.~Rempe}, and \bibinfo{author}{S.~Ritter}.
\newblock \emph{\bibinfo{title}{Frequency splitting of polarization eigenmodes
  in microscopic {F}abry-{P}erot cavities}}.
\newblock \bibinfo{journal}{New~J.~Phys.}, \textbf{\bibinfo{volume}{17}},
  \bibinfo{pages}{013053} (\bibinfo{year}{2015}).
\newblock \eprint{arXiv:1408.4367}.

\bibitem{Wang08}
\bibinfo{author}{X.~B. Wang}, \bibinfo{author}{J.~Q. You}, and
  \bibinfo{author}{F.~Nori}.
\newblock \emph{\bibinfo{title}{Quantum entanglement via two-qubit quantum
  {Z}eno dynamics}}.
\newblock \bibinfo{journal}{Phys.~Rev.~A}, \textbf{\bibinfo{volume}{77}},
  \bibinfo{pages}{062339} (\bibinfo{year}{2008}).

\bibitem{Maniscalco08}
\bibinfo{author}{S.~Maniscalco}, \bibinfo{author}{F.~Francica},
  \bibinfo{author}{R.~L. Zaffino}, \bibinfo{author}{N.~Lo~Gullo}, and
  \bibinfo{author}{F.~Plastina}.
\newblock \emph{\bibinfo{title}{Protecting Entanglement via the Quantum {Z}eno
  Effect}}.
\newblock \bibinfo{journal}{Phys.~Rev.~Lett.}, \textbf{\bibinfo{volume}{100}},
  \bibinfo{pages}{090503} (\bibinfo{year}{2008}).

\bibitem{Rossi09}
\bibinfo{author}{R.~Rossi}, \bibinfo{author}{K.~M. Fonseca~Romero}, and
  \bibinfo{author}{M.~C. Nemes}.
\newblock \emph{\bibinfo{title}{Semiclassical dynamics from {Z}eno-like
  measurements}}.
\newblock \bibinfo{journal}{Phys.~Lett.~A}, \textbf{\bibinfo{volume}{374}},
  \bibinfo{pages}{158} (\bibinfo{year}{2009}).

\bibitem{Shao09}
\bibinfo{author}{X.~Q. Shao}, \bibinfo{author}{L.~Chen},
  \bibinfo{author}{S.~Zhang}, and \bibinfo{author}{K.-H. Yeon}.
\newblock \emph{\bibinfo{title}{Fast CNOT gate via quantum {Z}eno dynamics}}.
\newblock \bibinfo{journal}{J.~Phys.~B}, \textbf{\bibinfo{volume}{42}},
  \bibinfo{pages}{165507} (\bibinfo{year}{2009}).

\bibitem{Chandrashekar10}
\bibinfo{author}{C.~M. Chandrashekar}.
\newblock \emph{\bibinfo{title}{{Z}eno subspace in quantum-walk dynamics}}.
\newblock \bibinfo{journal}{Phys.~Rev.~A}, \textbf{\bibinfo{volume}{82}},
  \bibinfo{pages}{052108} (\bibinfo{year}{2010}).

\bibitem{Shi12}
\bibinfo{author}{Z.~Shi}, \bibinfo{author}{Y.~Xia}, \bibinfo{author}{W.~H.Z.},
  and \bibinfo{author}{J.~Song}.
\newblock \emph{\bibinfo{title}{One-step preparation of three-particle
  Greenberger-Horne-Zeilinger state via quantum {Z}eno dynamics}}.
\newblock \bibinfo{journal}{Eur.~Phys.~J.~D}, \textbf{\bibinfo{volume}{66}},
  \bibinfo{pages}{127} (\bibinfo{year}{2012}).

\bibitem{Schleier10}
\bibinfo{author}{M.~H. Schleier-Smith}, \bibinfo{author}{I.~D. Leroux}, and
  \bibinfo{author}{V.~Vuleti\ifmmode~\acute{c}\else \'{c}\fi{}}.
\newblock \emph{\bibinfo{title}{States of an Ensemble of Two-Level Atoms with
  Reduced Quantum Uncertainty}}.
\newblock \bibinfo{journal}{Phys. Rev. Lett.}, \textbf{\bibinfo{volume}{104}},
  \bibinfo{pages}{073604}.
\newblock \doi{\bibinfo{doi}{10.1103/PhysRevLett.104.073604}}
  (\bibinfo{year}{2010}).
\newblock
  \urlprefix\url{http://link.aps.org/doi/10.1103/PhysRevLett.104.073604}.

\bibitem{Appel09}
\bibinfo{author}{J.~Appel}, \bibinfo{author}{P.~J. Windpassinger},
  \bibinfo{author}{D.~Oblak}, \bibinfo{author}{U.~B. Hoff},
  \bibinfo{author}{N.~Kj{\ae}rgaard}, and \bibinfo{author}{E.~S. Polzik}.
\newblock \emph{\bibinfo{title}{Mesoscopic atomic entanglement for precision
  measurements beyond the standard quantum limit}}.
\newblock \bibinfo{journal}{Proc.~Natl.~Acad.~Sci.~U.S.A.},
  \textbf{\bibinfo{volume}{106}}, \bibinfo{pages}{10960}
  (\bibinfo{year}{2009}).

\bibitem{Kuhr05}
\bibinfo{author}{S.~Kuhr}, \bibinfo{author}{W.~Alt},
  \bibinfo{author}{D.~Schrader}, \bibinfo{author}{I.~Dotsenko},
  \bibinfo{author}{Y.~Miroshnychenko}, \bibinfo{author}{A.~Rauschenbeutel}, and
  \bibinfo{author}{D.~Meschede}.
\newblock \emph{\bibinfo{title}{Analysis of dephasing mechanisms in a
  standing-wave dipole trap}}.
\newblock \bibinfo{journal}{Phys. Rev. A}, \textbf{\bibinfo{volume}{72}},
  \bibinfo{pages}{023406} (\bibinfo{year}{2005}).

\bibitem{Vandersypen05}
\bibinfo{author}{L.~M.~K. Vandersypen} and \bibinfo{author}{I.~L. Chuang}.
\newblock \emph{\bibinfo{title}{{NMR} techniques for quantum control and
  computation}}.
\newblock \bibinfo{journal}{Rev.~Mod.~Phys.}, \textbf{\bibinfo{volume}{76}},
  \bibinfo{pages}{1037} (\bibinfo{year}{2005}).

\bibitem{Gehr10}
\bibinfo{author}{R.~Gehr}, \bibinfo{author}{J.~Volz},
  \bibinfo{author}{G.~Dubois}, \bibinfo{author}{T.~Steinmetz},
  \bibinfo{author}{Y.~Colombe}, \bibinfo{author}{B.~L. Lev},
  \bibinfo{author}{R.~Long}, \bibinfo{author}{J.~Est\`eve}, and
  \bibinfo{author}{J.~Reichel}.
\newblock \emph{\bibinfo{title}{Cavity-Based Single Atom Preparation and
  High-Fidelity Hyperfine State Readout}}.
\newblock \bibinfo{journal}{Phys. Rev. Lett.}, \textbf{\bibinfo{volume}{104}},
  \bibinfo{pages}{203602}.
\newblock \doi{\bibinfo{doi}{10.1103/PhysRevLett.104.203602}}
  (\bibinfo{year}{2010}).

\bibitem{Rehacek01}
\bibinfo{author}{J.~Rehacek}, \bibinfo{author}{Z.~Hradil}, and
  \bibinfo{author}{M.~Jezek}.
\newblock \emph{\bibinfo{title}{Iterative algorithm for reconstruction of
  entangled states}}.
\newblock \bibinfo{journal}{Phys. Rev. A}, \textbf{\bibinfo{volume}{63}},
  \bibinfo{pages}{040303} (\bibinfo{year}{2001}).
\newblock \urlprefix\url{http://link.aps.org/doi/10.1103/PhysRevA.63.040303}.

\bibitem{Petz11}
\bibinfo{author}{D.~Petz} and \bibinfo{author}{C.~Ghinea}.
\newblock \emph{\bibinfo{title}{Introduction to quantum Fisher information}}.
\newblock In \emph{\bibinfo{booktitle}{Quantum Probability and Related
  Topics}}, volume~\bibinfo{volume}{1}, \bibinfo{pages}{261--281}
  (\bibinfo{year}{2011}).

\bibitem{Toth14}
\bibinfo{author}{G.~T{\'o}th} and \bibinfo{author}{I.~Apellaniz}.
\newblock \emph{\bibinfo{title}{{Quantum metrology from a quantum information
  science perspective}}}.
\newblock \bibinfo{journal}{Journal of Physics A: Mathematical and
  Theoretical}, \textbf{\bibinfo{volume}{47}}, \bibinfo{pages}{424006}
  (\bibinfo{year}{2014}).

\bibitem{Facchi09}
\bibinfo{author}{P.~Facchi}, \bibinfo{author}{G.~Marmo}, and
  \bibinfo{author}{S.~Pascazio}.
\newblock \emph{\bibinfo{title}{Quantum {Z}eno dynamics and quantum {Z}eno
  subspaces}}.
\newblock In \emph{\bibinfo{booktitle}{Journal of Physics: Conference Series}},
  volume \bibinfo{volume}{196}, \bibinfo{pages}{012017}.
  \bibinfo{organization}{IOP Publishing} (\bibinfo{year}{2009}).

\end{thebibliography}
}


{
\setlength{\parindent}{0pt}

\textbf{Acknowledgements:}
This work was supported by the European
Union Information and Communication Technologies project QIBEC
(Quantum Interferometry with Bose-Einstein Condensates) (GA~284584)
and by the Institut Francilien pour la Recherche sur les Atomes Froids
(IFRAF). We thank J{\"u}rgen Volz, Roger Gehr and Guilhem Dubois for
their contributions to early stages of this experiment.

\textbf{Author contributions:} 
G.~B. and L.~H. performed the
experiment; F.~H. contributed to setting it up. G.~B., L.~H.,
J.~E. and J.~R. contributed to data analysis and interpretation, as
well as to the manuscript.


\textbf{Supplementary Materials:}\\
Materials and Methods\\
Figs. S1 -- S7\\
Table S1\\
References (32--38)

\clearpage

\renewcommand{\figurename}{\textbf{Fig.}}
\renewcommand\thefigure{\textbf{S\arabic{figure}}}
\renewcommand{\tablename}{\textbf{Table}}
\renewcommand\thetable{\textbf{S\arabic{table}}}

\section*{Materials and Methods}
\vskip 1cm

\setcounter{figure}{0}

\subsection*{Atom chip and fiber Fabry-Perot cavity}

The experimental setup consists of an atom chip for the production of
ultracold trapped $^{87}$Rb atomic samples, integrated with a high
finesse fiber Fabry-Perot cavity. The experimental procedure to
produce the cold sample and transfer it to the optical trap in the
cavity has been described in \cite{Volz11}. The sample is loaded into
a single antinode of a 830\,nm standing wave dipole trap inside the
cavity. The {cavity has a length of $39\,\mu$m, estimated radii of
  curvature of the mirrors of $450\,\mu$m and $150\,\mu$m, and a
  finesse of 37000.} Parameters of the coupled atom-cavity system are
$(\kappa,\gamma,g)=2\pi\times(53,3,190)\,\mbox{MHz}$. In contrast to
previous experiments where the atomic cloud was placed at the
geometrical center of the cavity, here we load the atoms at the waist
of the 780\,nm probe laser which is located $8\,\mu\mbox{m}$ from the
input mirror (which has the larger radius of curvature).  This leads
to a $10\%$ enhancement of our atom-cavity coupling with respect to
previous experiments. Due to birefringence in the mirrors, our cavity
features two orthogonally polarised eigenmodes with a frequency
splitting of $540$\,MHz \cite{Volz11,Uphoff15}. In the present
experiments, probe and dipole beams are coupled to the
higher-frequency cavity mode. An external magnetic bias field of
12.5\,G is oriented parallel to this mode, thus both lasers are
$\pi$-polarised (Fig.~\ref{fig:levels}). To reduce the effect of
ambient magnetic fields, we implement our atomic qubit in the states
$\{\ket{0}\equiv |F=1,m_{F}=0\rangle,\ket{1}\equiv
|F=2,m_{F}=0\rangle\}$ which are insensitive to magnetic field
fluctuations. The probe laser is tuned close to resonance with the
$\ket{1}\rightarrow|F'=3,m_{F}=0\rangle$ transition of the D2 line of
$^{87}$Rb.

\begin{figure*}[hb]
\parbox[c]{.4\linewidth}{\includegraphics[width=\linewidth]{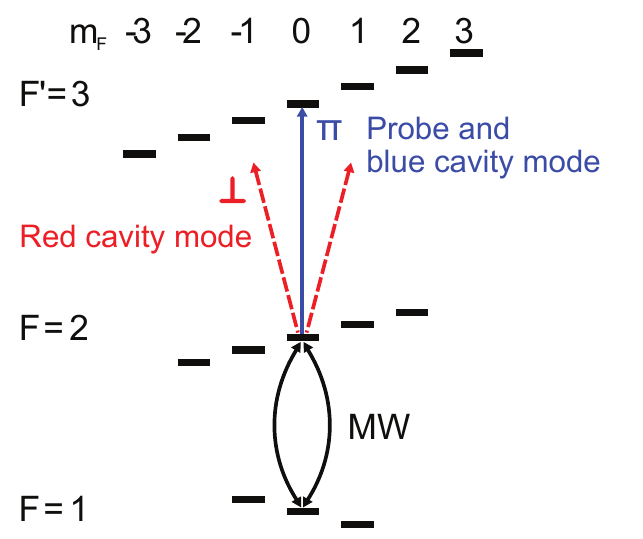}}\hfill%
\parbox[c]{.58\linewidth}{\caption{\small}\textbf{Relevant levels of $^{87}$Rb.} A magnetic field is applied
  along the polarization direction of the blue cavity mode (see
  text), which is also the polarization of the probe beam.}
\label{fig:levels}
\end{figure*}

The measurements performed in \cite{Haas14} indicated that the coherence
time of the final entangled state was mainly limited by the temperature
of the sample \cite{Kuhr05}. 
Due to an improved laser-cooling setup and optimisation of the cavity
locking apparatus, we now prepare atomic samples at a temperature of
$50\mu$K {(about 10 times colder than in \cite{Haas14})}, which
increases the coherence time by one order of magnitude. We expect the
temperature of the atomic ensemble to contribute only negligibly to
decoherence in the present experiment. Additionally, the delay between
the preparation of the entangled state and the tomography in the
present experiment is only $14\,\mu\mbox{s}$.

\subsection*{Atom number preparation and experimental sequence}

The atoms magnetically trapped by the atom chip are initially in the
$|F=2,m_{F}=2\rangle$ state. Once loaded in the cavity dipole trap, we
transfer them to $\ket{0}$ by means of three consecutive microwave
sweeps. In $F=1$, the atoms are not resonant with the light field in
the cavity, thereby acting as a dispersive medium with each atom
shifting the cavity resonance by $6$\,MHz. In order to prepare a
sample with precise atom number, we send probe light
into the cavity for a long time, causing
atom losses via light-assisted collisions (see \cite{Haas14}). By
simultaneously monitoring the cavity transmission, we can infer the
number of atoms left in the cavity and switch off the probe beam when
the desired value is reached. We terminate the preparation sequence
with a weak measurement pulse to check for
faulty sample preparation (atoms in $\ket{1}$ at $t=0$, $25\%$ of
runs).

The atom preparation process has a low probability to populate the
other Zeeman sub-levels of the $F=1$ manifold. {However, atoms in
  these sub-levels do not participate in any of the following steps;
  in particular, they do not contribute to the tomography measurement
  because they are not resonant with the microwave pulses.} We
verified the atom number in $\ket{0}$ after the preparation from a
separate tomography measurement of the initial state, as shown in the
upper panel of Fig.~2A. In the present experiments, we prepared a
target number of $40$ atoms in $F=1$ and measured the number of atoms
in the cavity that are in the correct Zeeman sub-state to be
$36\pm2$. This control measurement was repeated periodically to
  exclude drifts in the atom number.

\begin{figure*}
\centering
\includegraphics[width=0.7\textwidth]{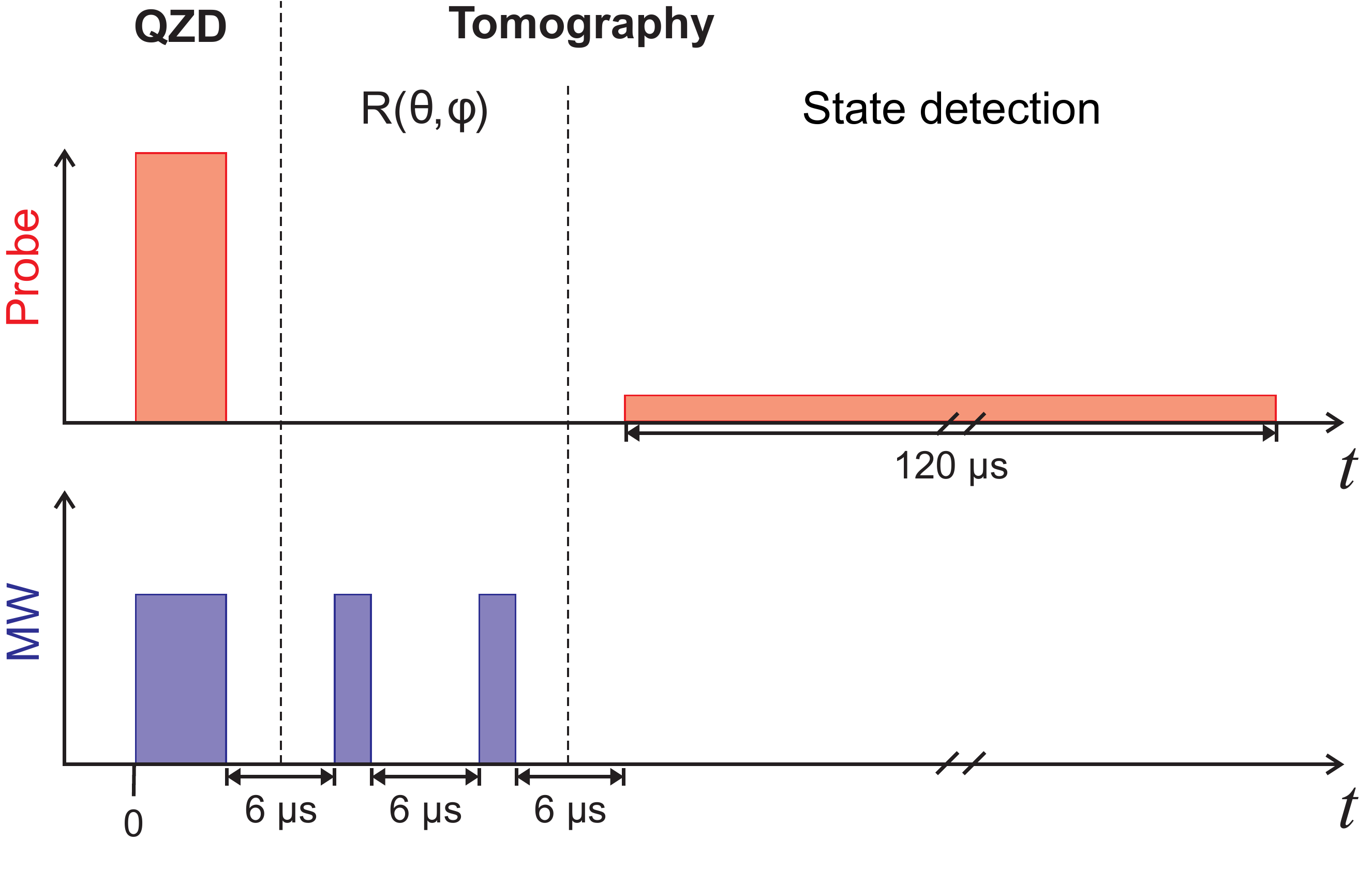}
\caption{\textbf{Timing sequence of QZD and tomography.} The
    coherent microwave pulse with simultaneous detection constitutes
    the actual quantum Zeno dynamics. Its duration is variable, but is
    typically close to the $\pi$-pulse duration $T=4.65\,\mu$s. Then
    follows the state tomography, which consists of a state rotation
    realized by two microwave pulses, and a detection pulse. The delay
    of $6\,\mu$s between any two pulses is imposed by hardware
    restrictions.}
\label{fig:sequence}
\end{figure*}

  Once the atoms are prepared in the desired state, we start one of
  the trajectories I and II, followed by state tomography, as shown in
  Fig.~\ref{fig:sequence}.  Initially, all the atoms are in $\ket{0}$
  so that the total spin points to the south pole of the Bloch sphere.
  The QZD sequence itself should start with a state fully inside
  $Z$. The most basic protocol would therefore consist in first
  applying a $\pi$ pulse to get from $\ket{0_{N}}$ to $\ket{1_{N}}$
  and then a $\theta=-\pi t/T$ pulse, accompanied by cavity
  measurement to perform the Zeno dynamics. However, it is
  crucial for the production of the $W$ state (Fig. 2A and 2B) to use
  a trajectory that passes exactly through the south pole. Therefore,
  to realize trajectory~I starting from $|0_{N}\rangle$, we first
  perform a $\theta=2\pi$ pulse, then a $-\pi$ pulse immediately
  followed by the $\theta=-\pi t/T$ pulse accompanied by
  measurement. It has been shown that this pulse sequence compensates
  for both pulse length and detuning fluctuations
  \cite{Vandersypen05}, ensuring that the trajectory in the absence of
  measurement passes exactly through the south pole of the Bloch
  sphere.
To perform trajectory~II, starting with all atoms in $\ket{0}$, we
first apply a small rotation around the $Y$ axis (for the experiment
shown in Fig.~2C $\phi=\pi/10$) and then a $\theta=\pi$ rotation
followed by the $\theta=\pi t/T$ rotation accompanied by the probing
of the cavity.

\subsection*{Cavity measurement}

The cavity measurement uses the method described previously in
\cite{Volz11,Haas14}. Note that it is robust with respect to
variations in the coupling rate: in order to keep the atoms
indistinguishable during measurement, the only requirement is to
realize high cooperativity $C\gg 1$ for each participating atom. Any
atom in that regime will cause the required binary cavity response --
full transmission for $\ket{0}$, full reflection for $\ket{1}$, and
thus, the measurement will not distinguish different atoms by their
coupling rates. The fulfillment of this requirement is experimentally
confirmed by the fact that we do not observe intermediate values of
transmission. (We can deliberately create them by strongly heating the
atom cloud.)

\subsubsection*{Tomographic measurement}

Immediately after the QZD we perform the 2D state tomography. Given
the atomic ensemble density matrix $\rho$, our tomography method
\cite{Haas14} measures the quantity
\begin{equation*}
  P_{0}(\theta,\phi)=\langle0_{N}|R^{\dagger}(\theta,\phi)\rho
  R(\theta,\phi)|0_{N}\rangle\,,
\end{equation*}
where $R(\theta,\phi)$ is a microwave
rotation and $\rho$ is the density matrix of the $N$ atomic qubits.
The dimension of the corresponding Hilbert space is $2^N$. However,
because $P_{0}(\theta,\phi)$ is always 0 for states outside the
  symmetric subspace (i.e., total spin $J<N/2$, see figure~1)
\cite{Haas14}, our measurement only yields information about the
symmetric part of the density matrix state $\rho_{s}=\Pi\rho\Pi$,
where $\Pi$ is the projector on the subspace of states with $J=N/2$
that are symmetric under particle exchange. The dimension of this
subspace is $N+1$. The Husimi-$Q$ distribution of $\rho_{s}$ is
defined as $Q(\theta,\phi)=\frac{N+1}{4\pi}P_0(\theta,\phi)$ and
contains all the information about $\rho_s$. In particular, the total
probability of being in the symmetric subspace $\Tr\rho_s$ is given by
the integral of $Q$.

For one complete 2D tomography measurement, we scan the tomography
angles $\theta$ (around $X$) and $\phi$ (around $Y$) each between
$-0.26\pi,...,0.26\pi$, resulting in a 7x7 grid of individual
measurements at different $(\theta,\phi)$.  To perform the required
rotation $R(\theta,\phi)$, we apply two consecutive microwave pulses,
with adjustable durations (0.4...1.2\,$\mu$s here) and phases; for
technical reasons, there is a delay of $6\,\mu\mbox{s}$ between
successive pulses leading to a total time delay of
$14\,\mu\mbox{s}$ between preparation and tomography. We then
probe the cavity during $120\,\mu\mbox{s}$ and measure the transmitted
and reflected photon counts on avalanche photodiodes (APDs), from which
we decide if the cavity transmission was high or low (see \cite{Gehr10} for more details). 
The experimental sequence is repeated in order to get at least 50 measurements for each
tomography angle, from which we deduce $P_{0}(\theta,\phi)$ as the frequency
of high transmission events.
The measurement of $P_{0}(\theta,\phi)$ is subject
to the detection errors $\epsilon_{01}$($\epsilon_{10}$) corresponding
to the probability to measure a high cavity transmission when there is in
reality at least one atom in $F=2$ (a low cavity transmission when in
reality all atoms are in $F=1$). The errors $\epsilon_{01}$ and
$\epsilon_{10}$ are computed as described in \cite{Haas14} and depend
on the average time for one atom to change its hyperfine state. These
lifetimes vary with the relative population of the two hyperfine
states which in turn depend on the tomography angle. We estimate the
lifetimes independently for each tomography angle to calculate
$\epsilon_{01}(\theta,\phi)$ and $\epsilon_{10}(\theta,\phi)$. In all
measurements they are below 6\%.

\subsection*{Maximum likelihood reconstruction of the density matrix}

Given our measurement results for $P_{0}(\theta,\phi)$, we reconstruct
the {symmetric part of the} density matrix, $\rho_s$, {including
  its off-diagonal elements}, in the Dicke state basis
$\{|n_N\rangle\}$. We use the maximum-likelihood algorithm described
in \cite{Rehacek01}. Starting from an arbitrary initial density
matrix, we iteratively find the matrix that maximizes the likelihood
of producing the observed results, taking into account the detection
errors $\epsilon_{01}(\theta,\phi)$ and $\epsilon_{10}(\theta,\phi)$
of the tomographic measurement. {We use the experimetally measured
  atom number N. Varying it on the scale of the experimental
  uncertainty produces no significant variation in the reconstruction
  results.}  Including too few basis states prevents the
reconstruction from correctly reproducing the experimental data while
including too many states increases the noise without giving
additional information. As can be seen in
Table~\ref{tab:SI_Table1_Reconstructed-populations-for-1}, the
reconstructed populations do not change when we include Dicke states
with $n>4$ and we therefore truncate the basis to the first fifth
Dicke states ($0\leq n \leq 4$) for all presented data.
Fig.~\ref{fig:SI_FIG2_Evolution-of-the} shows the evolution of all
five populations during QZD.  From the reconstructed $5 \times 5$
density matrix, we can calculate the Husimi-$Q$ function for arbitrary
angle and produce the high resolution density plots shown in Figure 2.

\begin{table}
\centering
\begin{tabular}{|c|c|c|c|c|c|c|c|c|}
\hline 
a)  & $\rho_{00}$  & $\rho_{11}$  & $\rho_{22}$  & $\rho_{33}$  & $\rho_{44}$  & $\rho_{55}$  & $\rho_{66}$  & $\Tr(\rho_{s})$\tabularnewline
\hline 
\hline 
$n_{\rm max}=2$  & 0.16  & 0.38  & 0.01  &  &  &  &  & 0.55\tabularnewline
\hline 
$n_{\rm max}=3$  & 0.17  & 0.36  & 0.01  & 0.01  &  &  &  & 0.55\tabularnewline
\hline 
$n_{\rm max}=4$  & 0.17  & 0.36  & 0.01  & 0.00  & 0.01  &  &  & 0.55\tabularnewline
\hline 
$n_{\rm max}=5$  & 0.17  & 0.36  & 0.01  & 0.00  & 0.00  & 0.00  &  & 0.55\tabularnewline
\hline 
$n_{\rm max}=6$  & 0.17  & 0.36  & 0.01  & 0.00  & 0.00  & 0.00  & 0.00  & 0.55\tabularnewline
\hline 
\end{tabular}

\centering
\begin{tabular}{|c|c|c|c|c|c|c|c|c|}
\hline 
b)  & $\rho_{00}$  & $\rho_{11}$  & $\rho_{22}$  & $\rho_{33}$  & $\rho_{44}$  & $\rho_{55}$  & $\rho_{66}$  & $\Tr(\rho_{s})$\tabularnewline
\hline 
\hline 
$n_{\rm max}=2$  & 0.03  & 0.13  & 0.22  &  &  &  &  & 0.38\tabularnewline
\hline 
$n_{\rm max}=3$  & 0.03  & 0.08  & 0.15  & 0.15  &  &  &  & 0.40\tabularnewline
\hline 
$n_{\rm max}=4$  & 0.02  & 0.07  & 0.11  & 0.12  & 0.08  &  &  & 0.41\tabularnewline
\hline 
$n_{\rm max}=5$  & 0.03  & 0.05  & 0.12  & 0.13  & 0.07  & 0.01  &  & 0.40\tabularnewline
\hline 
$n_{\rm max}=6$  & 0.02  & 0.07  & 0.12  & 0.11  & 0.07  & 0.01  & 0.01  & 0.41\tabularnewline
\hline 
\end{tabular}
\caption{\textbf{Reconstructed density matrix populations }$\rho_{ii}$ \textbf{for different basis truncations.}
The basis includes the Dicke states with $0,1,\ldots,n_{\rm max}$ atoms in $\ket{1}$.
(a) Reconstructed populations for the state in Fig.~2B obtained after a QZD of $t/T=0.96$.
(b) The same reconstruction but for the state displayed in the rightmost
frame of Fig. 2C. 
\label{tab:SI_Table1_Reconstructed-populations-for-1}}
\end{table}

To estimate the statistical error on the quantities that we infer from
the reconstructed density matrix, we use a bootstrapping method. We generate sets of artificial measurements
each having the same number of samples and the same average $P_{0}(\theta,\phi)$
as the experimental data. For each 2D tomography measurement, we then
operate the reconstruction algorithm on 1000 artificial datasets and calculate the quantity of interest
in order to obtain its standard deviation. We apply this method to obtain statistical error bars for the populations shown in figures 3,4 and for the quantum Fisher information. 

\begin{figure*}
\centering
\includegraphics[width=0.7\textwidth]{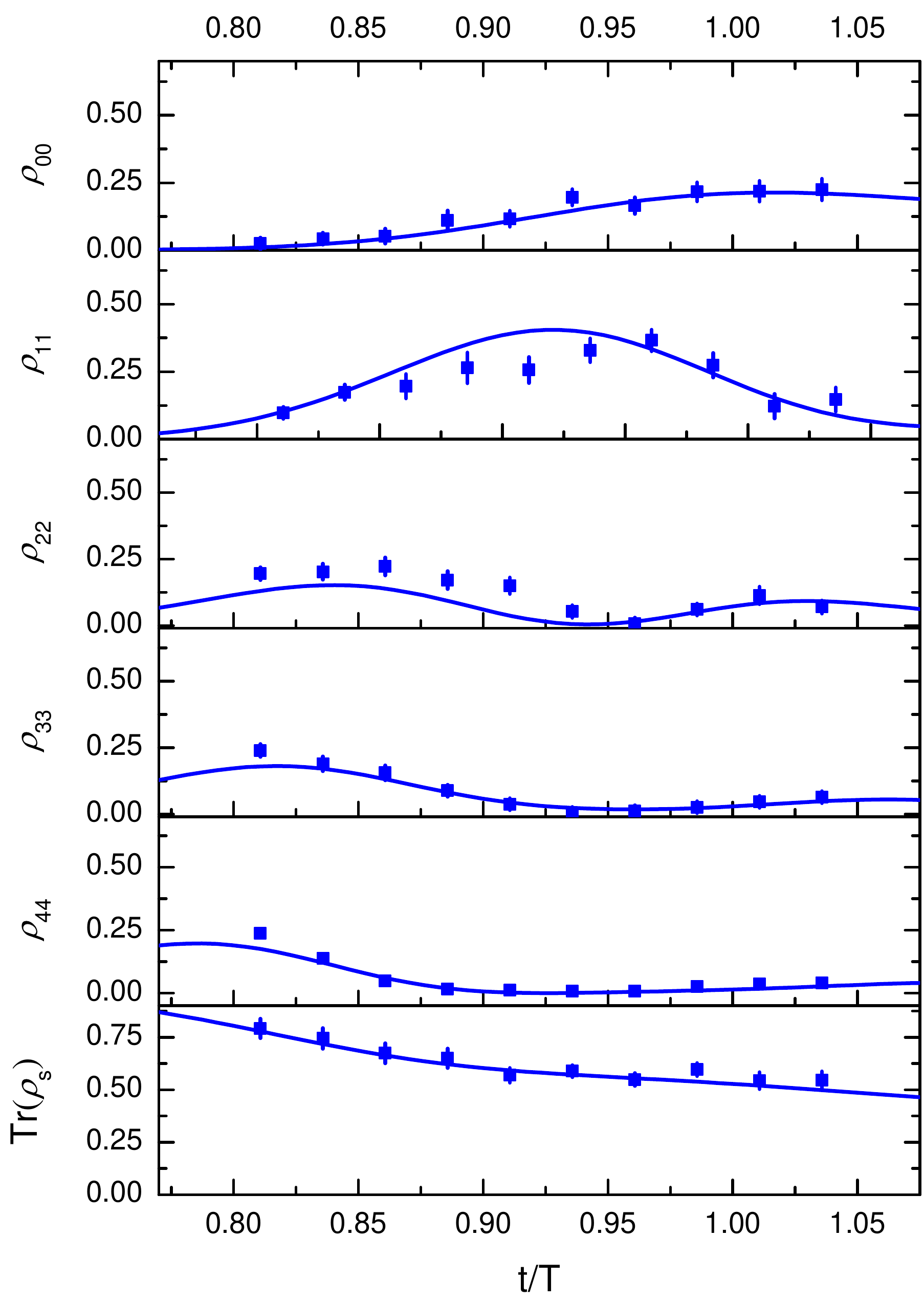} 
\caption{\textbf{Evolution of the populations during QZD along trajectory~I.} The points
  are the populations $\rho_{ii}$, with $i<5$, reconstructed from the
  tomography measurement shown in Fig. 2B. The lowest curve shows the
  total population in the symmetric subspace. The solid lines are the
prediction of Eq.~\ref{eq.mastereq_sp}.}
\label{fig:SI_FIG2_Evolution-of-the}
\end{figure*}

\subsection*{Criteria for multiparticle entanglement}

To assess the creation of multiparticle entanglement during QZD, we
use two different criteria for the states obtained following
trajectory~I and II. In the vicinity of the W-state (trajectory~I), we
employ the criterion described in \cite{Haas14} that gives a lower bound
on the entanglement depth of the atomic ensemble based on the
populations $\rho_{00}$ and $\rho_{11}$. For the states created along
trajectory~II, we compute a lower bound for the quantum Fisher
information $\Fq$. 

\subsubsection*{Establishing a lower bound on the quantum Fisher information}
We now show that $\Fq(\rho) \ge \Fq(\rho_{s})$, and that
computing the quantum Fisher information in the symmetric subspace
gives us a lower bound for $\Fq(\rho)$. We consider the completely
positive, trace preserving map $M$ which transforms $\rho$ into
$M[\rho]=\Pi \rho \Pi + \bar{\Pi} \rho \bar{\Pi}$, where 
$\Pi$ is defined as before and $\bar{\Pi} = 
\mathbf{1}
- \Pi$.  Under
a rotation of the atomic spin along the direction $\textbf{n}$ by an
angle $\theta$, $\rho$ transforms into $\rho(\theta )$. The quantum
Fisher information associated to the measurement of the angle
$\theta$, which we denote by $\Fq(\{ \rho(\theta) \})$, cannot
increase under the action of $M$ \cite{Petz11} and $\Fq(\{
\rho(\theta) \}) \geq \Fq(\{ M[ \rho(\theta) ] \})$.  Because the norm
of the atomic spin is conserved by rotations, we have $M[ \rho(\theta)
] = \rho_{s}(\theta) + \rho_{ns}(\theta)$, where $\rho_s(\theta)$ and
$\rho_{ns}(\theta)$ are respectively the rotated symmetric part of the
density matrix $\rho_s$ and the rotated non-symmetric part of the
density matrix $\rho_{ns} = \bar{\Pi} \rho \bar{\Pi}$.  We thus
conclude that
\begin{equation*}
\Fq(\{ \rho(\theta) \}) \ge \Fq(\{\rho_s(\theta) + \rho_{ns}(\theta)\}) = \Fq(\{\rho_s(\theta) \}) + \Fq(\{ \rho_{ns}(\theta)\}),
\end{equation*}
where the last equality comes from the additivity of the quantum Fisher information under direct sum \cite{Toth14}.

\subsubsection*{Computation of $\Fq$}
We compute $\Fq(\rho_{s})$ by numerically finding the optimal rotation
axis $\textbf{n}$ that maximizes the corresponding quantum Fisher
information $\Fq(\{ \rho_{s}(\theta) \})
=2\sum_{j,k}\frac{(p_{j}-p_{k})^{2}}{p_{j}+p_{k}}|\langle
j|J_{\textbf{n}}|k\rangle|^{2}$, where the $p_j$ and $|j\rangle$ are
the eigenvalues and eigenvectors of $ \rho_s$. {As a consistency
  check, we have also computed $\Fq(\rho_{s})$ for nominally
  nonentangled states in the experiment. We find
  $F_Q=38.7^{+6.8}_{-2.2}$ for the first state ($r_m=0$)
  in Fig.~4.}  In order to estimate our error on
$\Fq(\rho_{s})$, we use the bootstrapping method as previously
described. The errors indicated in the main text correspond to a
68\% confidence interval. The lower and upper errors are different
because the probability distribution that we deduce from the
bootstrapping method is slightly asymmetric. Extended Data
Fig.~\ref{fig:SI_Fisher} shows the evolution of $\Fq(\rho_{s})$ along
the trajectory~II with 68\% and 95\% confidence intervals. Note that
$\Fq(\rho_{s})$ is a very conservative estimate of the lower bound of
$\Fq(\rho)$ because it assumes $\Fq(\rho_{ns})=0$. Although we cannot
give a precise value to $\Fq(\rho_{ns})$ from our measurements, we can
estimate it to be on the order of $(N-n_{\rm sp})\Tr\rho_{ns}$. Here,
we have simply considered that the $n_{\rm sp}$ spontaneous emission
events that lead to a population $\Tr\rho_{ns}$ outside the symmetric
subspace correspond to a loss of $n_{\rm sp}$ atoms outside the qubit
subspace and that the remaining $N-n_{\rm sp}$ atoms are left in
$\ket{0}$. Including this contribution to our estimate of the lower
bound for $\Fq$ would increase it by approximately 25\%.

\begin{figure*}
\parbox[c]{.6\linewidth}{\includegraphics[width=\linewidth]{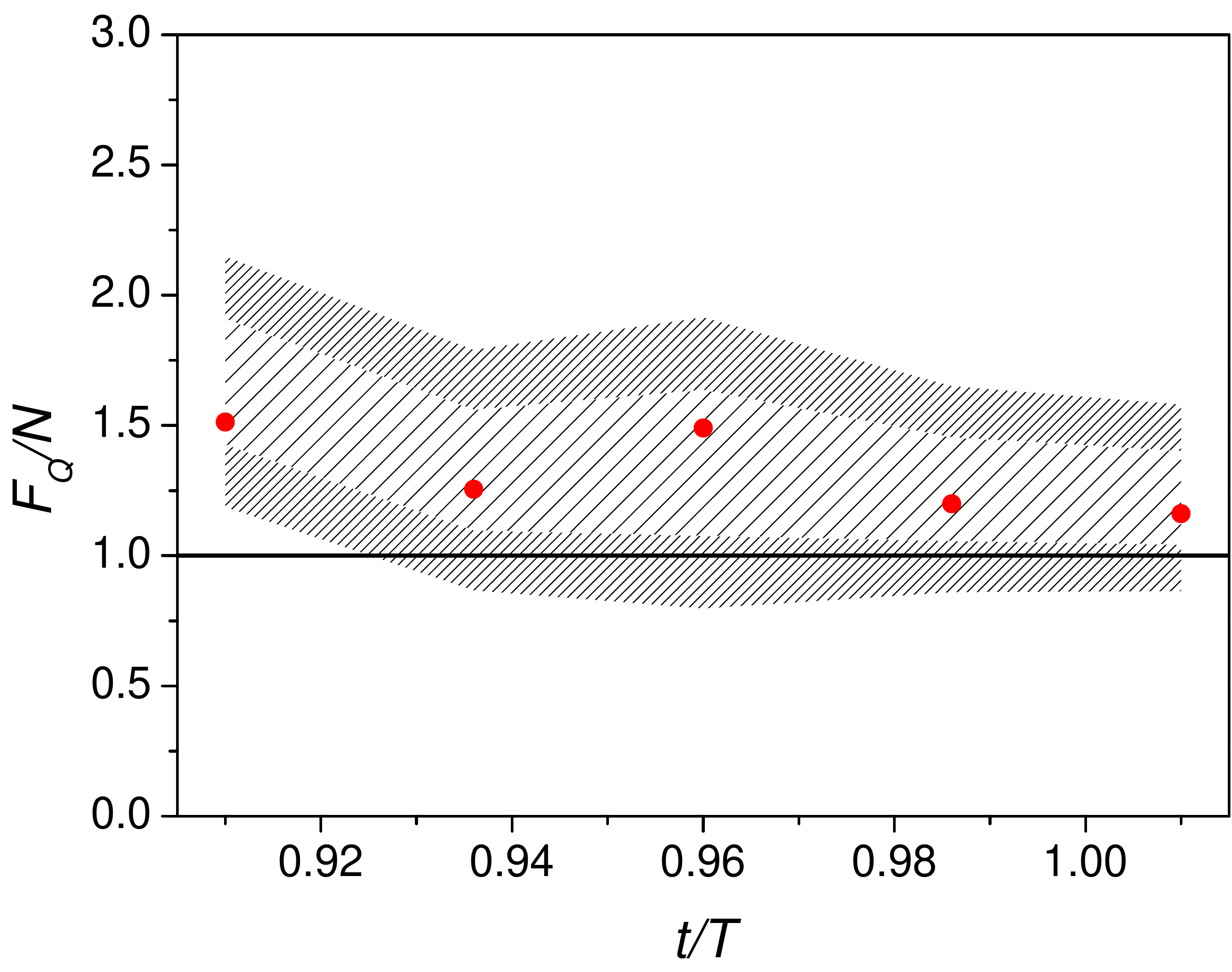}}\hfill%
\parbox[c]{.38\linewidth}{\caption{\textbf{Evolution of the quantum Fisher information during
    QZD along trajectory~II.}  We estimate a lower bound to the
  quantum Fisher information $\Fq$ of the atomic state by computing
  $\Fq$ of $\rho_s$ that we reconstruct from our tomography
  measurements.  A ratio $\Fq/N>1$ indicates an entangled state. The
  hatched areas are confidence intervals corresponding to 68\% and
  95\% probabilities.  The probability that the first point
  corresponds to an entangled state with $\Fq/N>1$ is above 99\%.  For
  the four other points, this probability is above 90\%.}
  \label{fig:SI_Fisher}}
\end{figure*}

\subsection*{Modeling the Quantum Zeno Dynamics}
Quantum Zeno Dynamics is most often described in terms of repeated
projective measurements that are regularly spaced in time with intervals of
unitary evolution in between. In our case, the measurement and the
unitary evolution are simultaneous and continuous processes.
As shown in \cite{Facchi09}, the expected dynamics are physically
equivalent in both cases and tend towards a unitary evolution confined
to the degenerate subspace of the measurement in the limit where
the measurement rate tends to infinity (\emph{cf.} dash-dotted lines
in Fig.~3). Because our measurement is continuous, a natural description
of our system is obtained following a master equation
approach. Without spontaneous emission, the
evolution of the atomic density matrix $\rho$ is given by 
\begin{equation}
\frac{d\rho}{dt}=\frac{1}{i\hbar}[H_{{\rm MW}},\rho]+d\rho d^{\dagger}-\frac{1}{2}\rho d^{\dagger}d-\frac{1}{2}d^{\dagger}d\rho\,.\tag{S1}
\label{eq.mastereq}
\end{equation}
The first term describes the unitary evolution due to the microwave
field with $H_{{\rm MW}}/\hbar=\Omega J_{x}$. The last three terms
account for the effect of the measurement by the cavity in the
Lindblad form.  Here we consider that light only enters the cavity
when the atoms are in $|0_{N}\rangle$, therefore the loss of a photon
from the cavity is described by a jump operator for the atoms that
projects them onto $|0_{N}\rangle$. The probability per unit time of
such a jump is $2\kappa\, n_{{\rm ph}}$ where $n_{{\rm
    ph}}=\Phi\sqrt{T_{0}}/\kappa$ is the average photon number in the
cavity when the atoms are in $|0_{N}\rangle$.  The expression of the
atom jump operator is thus
\begin{equation}
d=\sqrt{2\kappa\, n_{{\rm ph}}}|0_{N}\rangle\langle0_{N}|=\sqrt{r_{m}}|0_{N}\rangle\langle0_{N}|\,,\tag{S2}
\end{equation}
where $r_{m}\equiv 2\Phi\sqrt{T_0}$ is the effective
measurement rate of the cavity. The populations shown as dashed lines
in Fig.~2 are obtained by numerically solving $(\ref{eq.mastereq})$
for the photon flux used in the experiment. This simple model reproduces
well the evolution of the atomic state inside the symmetric subspace
but cannot account for the decay of the population in this subspace.

\subsubsection*{Effect of spontaneous emission}

\begin{figure}[tb]
\parbox[c]{.6\linewidth}{\includegraphics[width=\linewidth]{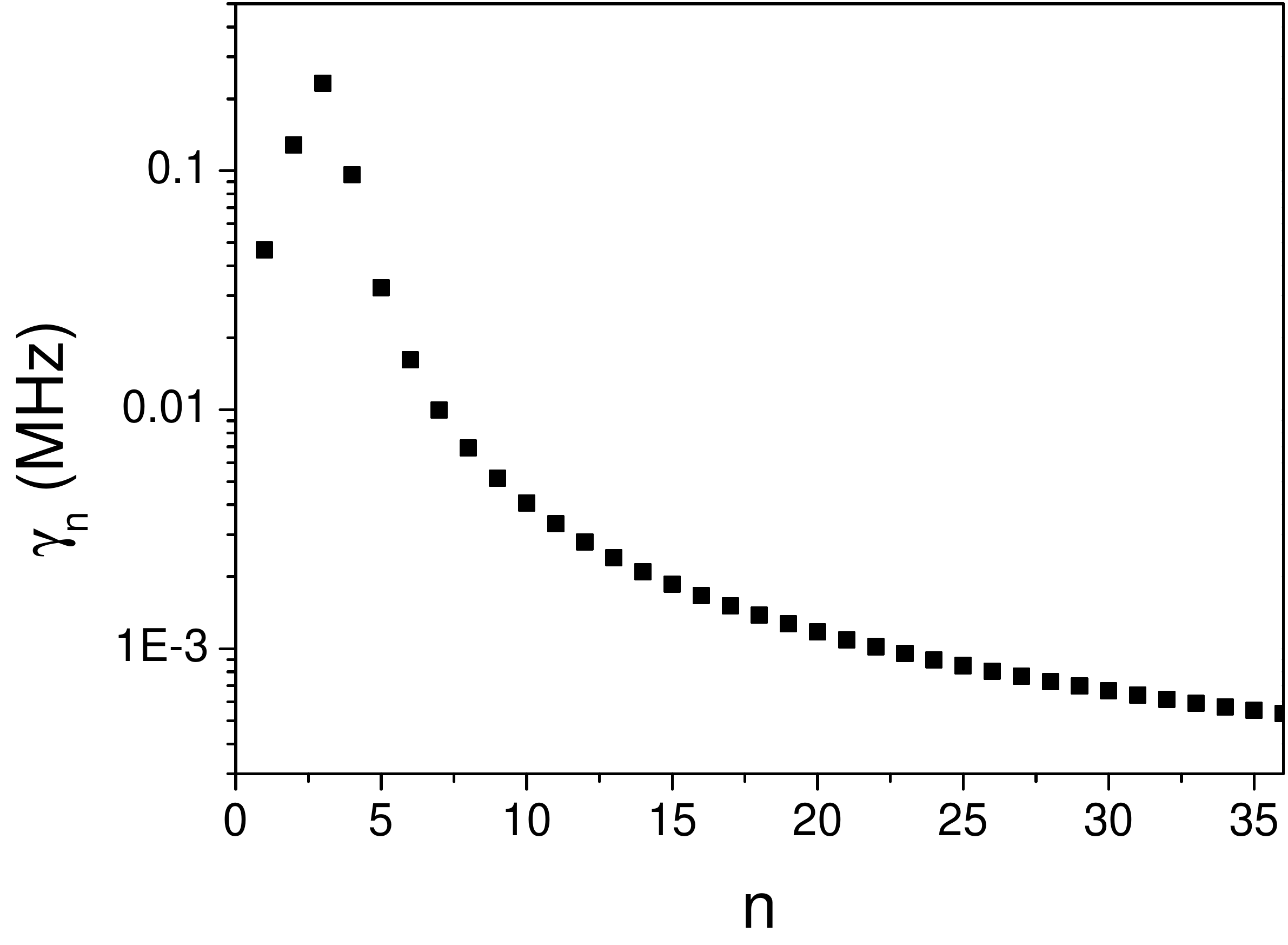}}\hfill%
\parbox[c]{.38\linewidth}{\caption{\small\textbf{Effective loss rate.}
    Calculated effective loss rates $\gamma_{n}$ when the atoms are in
    state $\ket{n_N}$.}
\label{fig:SI_Fig3_effectivegamma}}
\end{figure}

Spontaneous emission tends to populate undesired states outside
  the symmetric subspace. For example, when $\ket{1_N}$ is optically
  excited, it can emit to $\ket{J=N/2-1, n=1}$. In order to take
into account the effect of spontaneous emission and calculate the
expected decay of the population in the symmetric subspace, we
consider an anti-Hermitian operator $\bar{H}_{{\rm loss}}$ and the
modified master equation
\begin{equation}
\frac{d\rho}{dt}=\frac{1}{i\hbar}[H_{{\rm
    MW}},\rho]+\frac{1}{i\hbar}\{\bar{H}_{{\rm loss}},\rho\}+d\rho
d^{\dagger}-\frac{1}{2}\rho
d^{\dagger}d-\frac{1}{2}d^{\dagger}d\rho\,.\tag{S3} \label{eq.mastereq_sp}
\end{equation}
We suppose that $\bar{H}_{{\rm loss}}$ is diagonal in the Dicke state
basis with matrix elements given by $\langle n_{N}|\bar{H}_{{\rm loss}}|n_{N}\rangle=-i\gamma_{n}$
where $\gamma_{n}=\gamma\, p_{e,n}$ and $p_{e,n}$ is the probability
for one atom among the $n$ atoms in $\ket{1}$ to be optically
exited. Here, we neglect the probability ($\sim1/N$) to fall back
into the symmetric subspace after a spontaneous emission. The probability
of excitation $p_{e,n}$ is calculated independently from the steady
state solution of the atom-cavity master equation describing the coupling
of $n$ atoms to the TEM$_{00}$ mode taking into account all known
experimental parameters for our setup (magnetic field, lattice depth,
...) as in \cite{Volz11}, and in particular the presence of the orthogonally
polarized TEM$_{00}$ cavity mode. Fig.~\ref{fig:SI_Fig3_effectivegamma}
shows the calculated rates $\gamma_{n}$ as a function of $n$ corresponding
to the photon flux used to measure the data shown in Fig.~2. 
We then numerically solve (\ref{eq.mastereq_sp}) and obtain the populations
shown as solid lines in Figs.~3 and 4. The calculated
number of spontaneous emission events during the QZD explains the observed decay of the population $\Tr\rho_s$ in the symmetric subspace. Fig.~\ref{fig:SI_LongTimesQZD} shows 1D tomography measurements for evolution times spanning a larger interval than in the main text. The observed reduction in amplitude of the measured curves
is also well explained by our model. Finally,
Fig.~\ref{fig:SI_HusimiQ} shows the expected Husimi-$Q$ distributions
obtained from the model and compares them to the ones computed from
the reconstructed density matrices. 

\begin{figure*}
\centering
\includegraphics[width=0.8\textwidth]{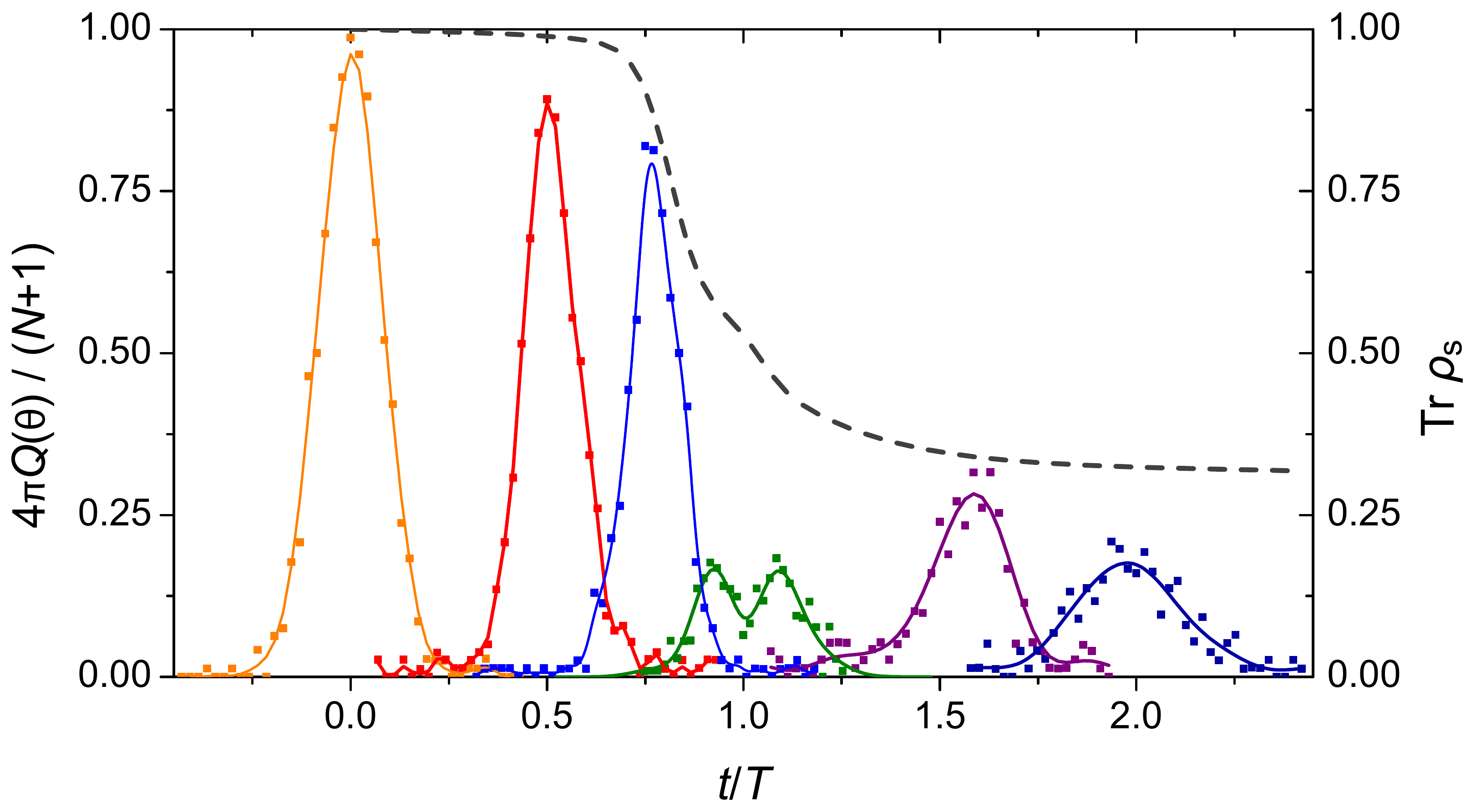}
\caption{\textbf{Long time evolution of the Husimi-$Q$ distribution
    during QZD.} Each colored curve corresponds to a 1D measurement of
  the Husimi-$Q$ distribution for a different evolution time along
  trajectory~I. Solid lines are guides to the eye.  For times shorter
  than the $\pi$ pulse time (orange, red, blue data), we observe the
  rotation of the initial coherent state on the Bloch sphere. In the
  vicinity of the $\pi$ pulse (green data), the state is strongly
  modified due to QZD. For larger times (purple, dark blue data), the
  state recovers a gaussian like shape, but with a reduced amplitude
  compared to an ideal coherent state. This is due to the effect of
  spontaneous emission, which leads to a decrease of the population in
  the symmetric subspace. The dashed gray line shows the evolution of
  the symmetric subspace population expected from the full model.}
\label{fig:SI_LongTimesQZD}
\end{figure*}


\begin{figure*}
\centering
\includegraphics[width=\textwidth]{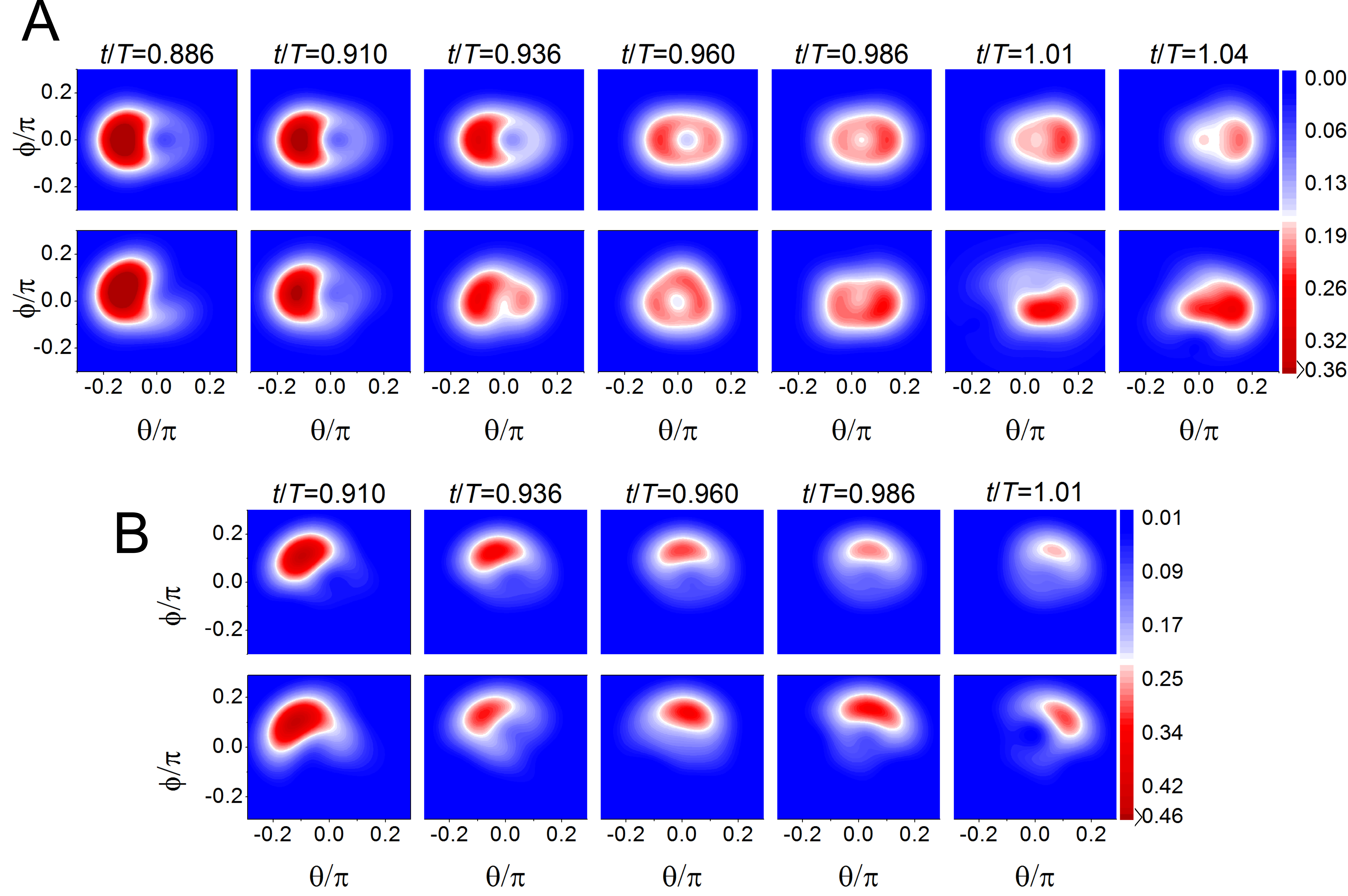} 
\caption{\textbf{Predictions of the full model.} Comparison of the
  predicted Husimi-Q distributions calculated from the full model
  including spontaneous emission (upper rows) to the
  experimentally obtained distributions 
  (lower rows, same as in Fig.~2 in the main text). Panel A (B)
  corresponds to the trajectory~I (II).}
\label{fig:SI_HusimiQ}
\end{figure*}

\end{document}